\documentclass[preprint2]{aastex}

\slugcomment{Version: \today}

\usepackage{subfigure}
\usepackage{float}
\usepackage{color}
\usepackage{hyperref}

\def\GIII      {G043.16$+$0.01}
\def\Jfive     {J1905$+$0952}
\def\Jtwo      {J1922$+$0841}

\def\GVIII     {G048.60$+$0.02}
\def\Jseven    {J1917$+$1405}
\def\Jthree    {J1913$+$1307}
\def\Jfour     {J1924$+$1540}

\def\uas       {$\mu$as}
\def\deg       {$^\circ$}
\def\kms       {km~s$^{-1}$}
\def\masy      {mas~yr$^{-1}$}
\def\jybeam    {Jy~beam$^{-1}$}
\def\mjybeam   {mJy~beam$^{-1}$}

\def\VLSR  {$V_{\rm LSR}$}

\def\to    {$-$}
\def\hho   {H$_2$O}
\def\meth  {CH$_3$OH}
\def\AIPS  {$\mathcal{AIPS}$}

\def\c     {$\sqrt{}$}
\def\x     {\phantom{$\sqrt{}$}}
\def\p     {\phantom{0}}

\def\mux   {\ifmmode {\mu_x}\else {$\mu_x$}\fi}
\def\muy   {\ifmmode {\mu_y}\else {$\mu_y$}\fi}
\def\mura  {\ifmmode {\mu_{\alpha}}\else {$\mu_{\alpha}$}\fi}
\def\mude  {\ifmmode {\mu_{\delta}}\else {$\mu_{\delta}$}\fi}

\def\dx    {$\Delta x$}
\def\dy    {$\Delta y$}

\def\Ts    {\ifmmode{\Theta_s}\else{$\Theta_s$}\fi}
\def\Tdot  {\ifmmode{d\Theta\over dR}\else{$d\Theta\over dR$}\fi}
\def\Rs    {\ifmmode{R_s}\else{$R_s$}\fi}
\def\To    {\ifmmode{\Theta_0}\else{$\Theta_0$}\fi}
\def\Ro    {\ifmmode{R_0}\else{$R_0$}\fi}

\def\Vo    {\ifmmode {V^{Std}_\odot}\else {$V^{Std}_\odot$}\fi}
\def\Uo    {\ifmmode {U^{Std}_\odot}\else {$U^{Std}_\odot$}\fi}
\def\Wo    {\ifmmode {W^{Std}_\odot}\else {$W^{Std}_\odot$}\fi}
\def\VH    {\ifmmode {V^H_\odot}\else {$V^H_\odot$}\fi}
\def\UH    {\ifmmode {U^H_\odot}\else {$U^H_\odot$}\fi}
\def\WH    {\ifmmode {W^H_\odot}\else {$W^H_\odot$}\fi}
\def\V     {\ifmmode {V_\odot}\else {$V_\odot$}\fi}
\def\U     {\ifmmode {U_\odot}\else {$U_\odot$}\fi}
\def\W     {\ifmmode {W_\odot}\else {$W_\odot$}\fi}
\def\VGC   {\ifmmode {V_\odot^{GC}}\else {$V_\odot^{GC}$}\fi}
\def\UGC   {\ifmmode {U_\odot^{GC}}\else {$U_\odot^{GC}$}\fi}
\def\WGC   {\ifmmode {W_\odot^{GC}}\else {$W_\odot^{GC}$}\fi}

\def\Vs    {\ifmmode {V_s}\else {$V_s$}\fi}
\def\Us    {\ifmmode {U_s}\else {$U_s$}\fi}
\def\Ws    {\ifmmode {W_s}\else {$W_s$}\fi}

\def\Vsbar {\ifmmode {\overline{V_s}}\else {$\overline{V_s}$}\fi}
\def\Usbar {\ifmmode {\overline{U_s}}\else {$\overline{U_s}$}\fi}
\def\Wsbar {\ifmmode {\overline{W_s}}\else {$\overline{W_s}$}\fi}

\usepackage{ulem}

\def\coIII {$^{13}$CO}

\begin{document}

\title{
Parallaxes for W49N and \GVIII:\\
Distant Star Forming Regions in the Perseus Spiral Arm
}

\author{
 B. Zhang\altaffilmark{1,2},
 M. J. Reid\altaffilmark{3},
 K. M. Menten\altaffilmark{1},
 X. W. Zheng\altaffilmark{4},
 A. Brunthaler\altaffilmark{1},
 T. M. Dame\altaffilmark{3},
 Y. Xu\altaffilmark{5}
}

\altaffiltext{1}{Max-Plank-Institut f\"ur Radioastronomie, Auf dem H\"ugel
69, 53121 Bonn, Germany}

\altaffiltext{2}{Shanghai Astronomical Observatory, Chinese Academy of
Sciences, Shanghai 200030, China}

\altaffiltext{3}{Harvard-Smithsonian Center for Astrophysics, 60
 Garden Street, Cambridge, MA 02138, USA}

\altaffiltext{4}{Department of Astronomy, Nanjing University, Nanjing
    210093, China}

\altaffiltext{5}{Purple Mountain Observatory, Chinese Academy of
Sciences, Nanjing 210008, China}

\begin{abstract}

We report trigonometric parallax measurements of 22 GHz \hho\ masers in
two massive star-forming regions from VLBA observations as part of the
BeSSeL Survey.  The distances of $11.11^{+0.79}_{-0.69}$~kpc to W49N
(\GIII) and $10.75^{+0.61}_{-0.55}$~kpc to \GVIII\ locate them in a
distant section of the Perseus arm near the solar circle in the first
Galactic quadrant.  This allows us to locate accurately the inner
portion of the Perseus arm for the first time.  Combining the present
results with sources measured in the outer portion of the arm in the
second and third quadrants yields a global pitch angle of 9.5\deg\ $\pm$
1.3\deg\ for the Perseus arm.  We have found almost no \hho\ maser
sources in the Perseus arm for 50\deg\ $< \ell <$ 80\deg, suggesting
that this $\approx 6$ kpc section of the arm has little massive star
formation activity.

\end{abstract}


\maketitle

\section{INTRODUCTION}
\label{sec:intro}

While the nature and even the number of spiral arms in the Milky Way is
still debated, mounting evidence suggests that the Perseus arm is one of
two major spiral
arms~\citep{2000A&A...358L..13D,2005ApJ...630L.149B,2009PASP..121..213C}.
It may emerge from the far end of the bar and wrap through the inner
Galaxy in the first quadrant (inner portion of the Perseus arm) and the
outer Galaxy in the second and third quadrants (outer portion of the
Perseus arm).  Since the inner portion of the Perseus arm lies at a
smaller Galactic radius and closer to the bar than the outer portion of
the arm, one would expect it to be more prominent in molecular gas and
star formation, yet very little is known about it owing to its great
distance and its kinematic blending with nearer material in the inner
Galaxy.

Recent improvements in radio astrometry with Very Long Baseline
Interferometry (VLBI) techniques have yielded parallaxes and proper
motions to star-forming regions in the Galaxy with accuracies of $\sim$
10 \uas\ and $\sim$ 0.1 \masy, respectively
(e.g.~\citealt{2009ApJ...700..137R,2012PASJ...64..136H}).  Parallax
measurements for a reasonably dense sampling of sources in spiral arms
will help us to fully trace the spiral structure of the Galaxy. To reach
this goal, we are using the NRAO\footnote{The National Radio Astronomy
Observatory is a facility of the National Science Foundation operated
under cooperative agreement by Associated Universities, Inc} Very Long
Baseline Array (VLBA) to conduct a key science project, the BeSSeL (Bar
and Spiral Structure Legacy)
Survey\footnote{\url{http://bessel.vlbi-astrometry.org/}}, to study the
structure and kinematics of the Galaxy by measuring trigonometric
parallaxes and proper motions for hundreds of 22 GHz \hho\ and 6.7/12.2
GHz \meth\ maser sources associated with massive star-forming regions.

In this paper, we present trigonometric parallax measurements of 22 GHz
\hho\ masers toward two high-mass star-forming regions, W49N (\GIII) and
\GVIII, in the inner portion of the Perseus arm.

\section{OBSERVATIONS AND CALIBRATION PROCEDURES}
\label{sec:obs}

Our observations of 22 GHz \hho\ masers toward \GIII\ and \GVIII,
together with several compact extragalactic radio sources, were carried
out under VLBA program BR145B with 12 epochs spanning about one year.
For these sources, the parallax signature in Declination was
considerably smaller than for Right Ascension, and we scheduled the
observations so as to maximize the Right Ascension parallax offsets as
well as to minimize correlations among the parallax and proper motion
parameters.  The observations near each Right Ascension parallax peak
were grouped as listed in Table~\ref{tab:obs}.  At each epoch, the
observations consisted of four 0.5-hour ``geodetic blocks'' (used to
calibrate and remove unmodeled atmospheric signal delays), with three
1.7-hour periods of phase-referenced observations inserted between the
blocks.  In the phase-referenced observations, we cycled between the
target maser and several background sources, switching sources every 20
to 30 seconds.  Table~\ref{tab:src} lists the observed source positions,
intensities, source separations, reference maser radial velocities and
restoring beams.  The typical on-source integration time per epoch for
the maser source and each background source were 0.64 and 0.30 hour for
\GIII, and 0.79 and 0.25 hour for \GVIII, respectively.

Our general observing setup and calibration procedures are described in
\citet{2009ApJ...693..397R}; here we discuss only aspects of the
observations that are specific to the maser sources presented in this
paper.  We used four adjacent intermediate frequency (IF) bands
with 16 MHz, each in both right and left circular polarization (RCP and
LCP); the second band contained the maser signals, the center
\VLSR\ is 10 and 26 \kms\ for \GIII\ and \GVIII, respectively.  The
spectral-channel spacing was 31.25 kHz corresponding to 0.42 \kms\ in
velocity.  We observed three International Celestial Reference Frame
(ICRF) sources: 3C345 (J2253+1608), 3C454 (J1642+3948) and
J1925+2106~\citep{1998AJ....116..516M}, near the beginning, middle and
end of the phase-referencing observations in order to monitor delay and
electronic phase differences among the observing bands.  The data
correlation was performed with the DiFX\footnote{DiFX: A software
Correlator for VLBI using Multiprocessor Computing Environments, is
developed as part of the Australian Major National Research Facilities
Programme by the Swinburne University of Technology and operated under
licence} software correlator \citep{2007PASP..119..318D} in Socorro, NM.
The data reduction was conducted using the NRAO's Astronomical Image
Processing System (\AIPS) together with scripts written in ParselTongue
\citep{2006ASPC..351..497K}.
Since in our case the masers are much stronger than the background
sources, we used a spectral channel with strong and relatively compact
maser emission as the interferometer phase reference.  This is necessary
to extend the coherence time of the interferometer and allow all data to
be used to make an image.  This allows us to detect weak background
sources and other maser spots in many spectral channels in order to
determine their positions respect to the reference maser spot.  When
differencing the positions of the other maser spots and the background
sources, structure in the reference spot cancels and as such does not
affect parallax measurements.
After we performed the calibration for the polarized bands separately,
we combined the RCP and LCP bands to form Stokes I and imaged the
continuum emission of the background sources from the four frequency
bands simultaneously using the \AIPS\ task {\it IMAGR}. For the masers,
we also formed Stokes I and then imaged the emission in each spectral
channel. Then, we fitted elliptical Gaussian brightness distributions to
the images of strong maser spots and the background sources using the
\AIPS\ task {\it SAD} or {\it JMFIT}.

\section{ASTROMETRIC PROCEDURES}
\label{sec:procedure}


Data used for parallax and proper motion fits were residual position
differences between maser spots and background sources in eastward (\dx\
= $\Delta\alpha\cos\delta$) and northward (\dy\ = $\Delta\delta$)
directions.  These residual position differences are relative to the
coordinates used to correlate the VLBA data and shifts applied in
calibration.   The data were modeled by the parallax sinusoid in both
coordinates (determined by a single parameter, the parallax) and a
linear proper motion in each coordinate.  Because systematic errors
(owing to small uncompensated atmospheric delays and, in some cases,
varying maser and calibrator source structures) typically dominate over
thermal noise when measuring relative source positions, we added ``error
floors'' in quadrature to the formal position uncertainties. We used
different error floors for the \dx\ and \dy\ data and adjusted them to
yield post-fit residuals with reduced $\chi^2$ near unity for both
coordinates.

As discussed in \citet{2012ApJ...744...23Z}, the apparent motions of the
maser spots can be complicated by a combination of spectral and spatial
blending and changes in intensity. Thus, for parallax fitting, one needs
to find stable, unblended spots and preferably use many maser
spots to average out these effects.  We selected maser spots
brighter than 50 and 1 \jybeam, which are $\approx$ 1/200 of the peak
brightness, for \GIII\ and \GVIII, respectively. We considered maser
spots at different epochs as being from the same feature if their
separation from the middle-epoch position was less than 5 mas (a $<$ 5
mas shift over 0.5 years corresponds to a maser spot motion of $<$ 260
\kms\ at a distance of 11 kpc).

\hho\ masers can be time-variable with lifetimes of months to
years. For solid parallax fits, we selected only maser spots persisting
over all epochs to avoid large correlations between parallax and proper
motion.  We first fitted a parallax and proper motion to each \hho\
maser spot relative to each background source separately.
Since one expects the same parallax for all maser spots, we did a
combined solution (fitting with a single parallax parameter for all
maser spots, but allowing for different proper motions for each maser
spot) using all maser spots and background sources.


In general, \hho\ maser spots usually are not distributed uniformly
around the exciting star and their kinematics can be complicated by a
combination of expansion and rotation \citep{1992ApJ...393..149G}; this
limits the accuracy of estimates of the absolute proper motion of the
exciting star(s).  Therefore, we modeled the {\it relative} motions of
maser spots distributed across the source in order to solve for the
motion of the phase-reference spot relative to the central star.  Owing
to the large field of view of the maser spots, especially in \GIII, for
measuring relative motions we used only the inner-five antennas of the
VLBA (to allow a wide field of view and to limit the number of pixels
needed to map the sources).


In order to model the expansion and rotation of the entire maser source,
we adopted a Bayesian fitting procedure using a Markov chain Monte Carlo
method to explore parameter space, assuming the probability distribution
for the data uncertainties follows the ``Conservative Formulation'' of
\citet{2006OUP.book......S}, which does not have a large penalty for
out-lying data points (i.e., maser spots whose motions do not follow the
simple expanding model).  The details on the maser kinematics model
and the Bayesian fitting procedure are described in
\citet{1992ApJ...393..149G} and \citep{2010ApJ...720.1055S},
respectively. We first used a simple radial expanding outflow model
(model A) and then an expanding outflow with rotation (model B).  The
global parameters we estimated include the position $(x_0,y_0)$ and
motion ($V_{0x}$, $V_{0y}$) of the expansion center (relative to the
reference maser spot); the \VLSR\ of the expansion center $V_{0r}$; an
expansion speed $V_{\rm exp}$ at 1\arcsec\ radius from the expansion
center and the exponent $\gamma$ that allows for acceleration as a
power law for velocity as a function of distance from the
expansion center; and the rotation speed at 1\arcsec\ radius from the
spin axis with two orientation angles of $\theta$ and $\phi$, the
azimuth and elevation of the spin axis in the reference frame of the
model, respectively.  In addition to the global parameters, for
each maser spot we estimated its offset along the line of sight from the
reference maser spot.

\section{RESULTS}
\label{sec:results}

\subsection{W49N}

\begin{figure}[H]
  \centering
  \includegraphics[angle=0,scale=0.75]{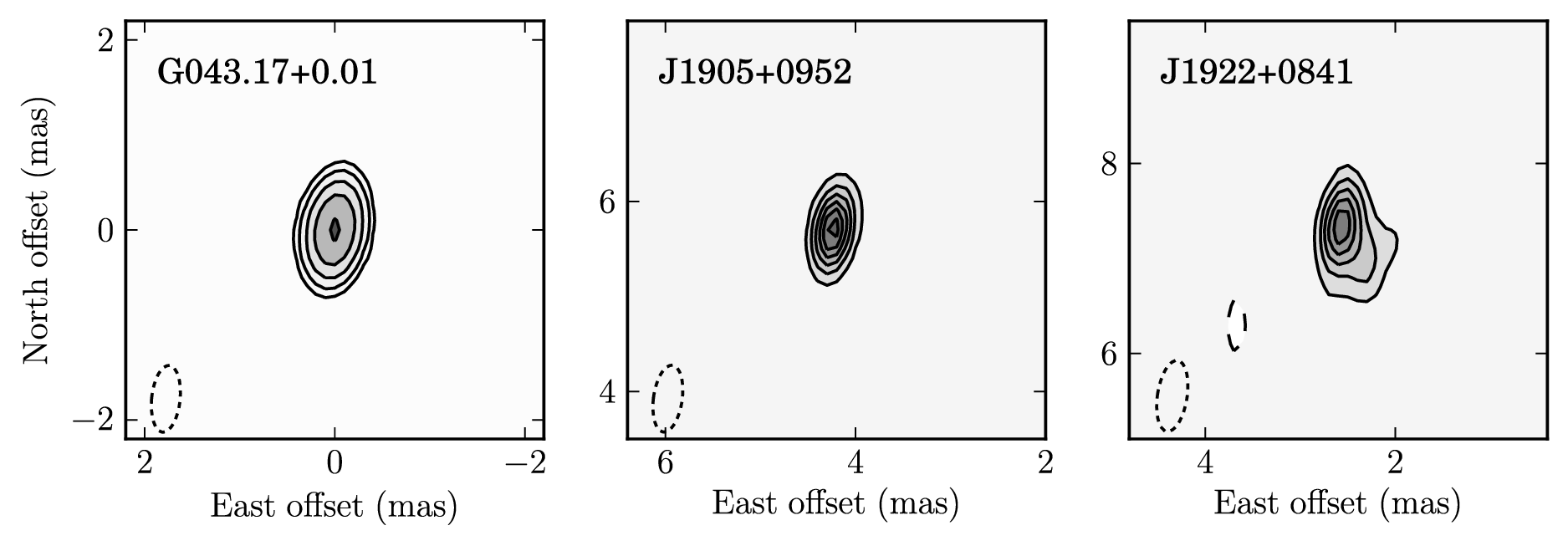}
  \caption{
  Images of the \hho\ maser reference spot at \VLSR\ of --4.75 \kms\ and
  the extragalactic radio sources used for the parallax measurements of
  W49N (\GIII) at the last epoch (2011 April 18).  Source names are in the upper left
  corner and the restoring beam (dotted ellipse) is in the lower left
  corner of each panel.  Contour levels for \GIII\ maser emission are 250
  \jybeam\ $\times 2^n$, $n$ = 0 \ldots 5, and for the background sources
  (\Jfive\ and \Jtwo) are spaced linearly at 0.02 and 0.20 \mjybeam,
  respectively.
  \label{fig:g043_mq}}
\end{figure}

W49N (\GIII) is a complex region of recent star formation containing the
most luminous \hho\ maser site in the Galaxy
\citep{1969Natur.221..626C,1970ApJ...160L..63B}.  For the parallax
measurement of W49N we phase-referenced to the maser spot at \VLSR\ of
--4.75 \kms.  Both background sources were detected at all epochs,
except for \Jtwo\ at the second epoch.   Figure~\ref{fig:g043_mq} shows
images of the reference maser spot and the two background sources at
the last epoch (BR145BC). 

\begin{figure}[H]
  \centering
  \includegraphics[angle=0,scale=0.75]{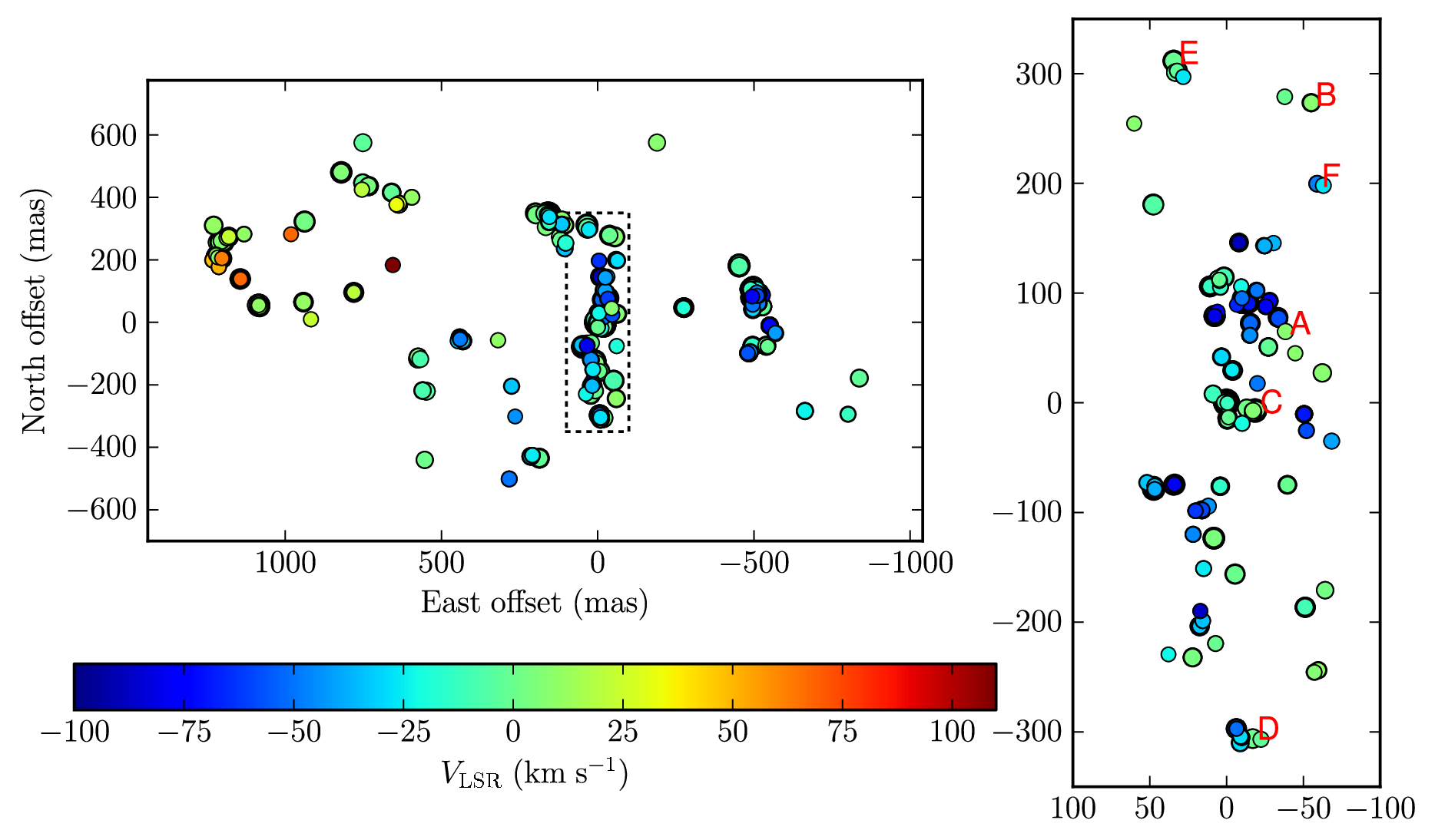}
  \caption{
  {\it Left panel}: spatial distribution of the \hho\ maser spots (with
  brightness $\ge$ 50 \jybeam\ for a restoring beam of $\sim 3 \times 1$
  mas) toward W49N from VLBA observations at the last epoch (2011 April 18).
  Each maser spot is represented by a filled circle whose area is
  proportional to the logarithm of the flux density.  {\it Right panel}:
  blow-ups of maser spots (with brightness $\ge$ 50 \jybeam\ for a
  restoring beam of $\sim 0.8 \times 0.4$ mas) distribution in the field
  indicated with a dotted box in {\it left panel}.  All the maser spots
  used for the parallax fit are from this field.  Each region including
  maser feature used for parallax fit is labeled with a letter.  The
  reference maser spot is located in region C.  The color bar denotes
  the \VLSR\ range from --100 to +110 \kms. \newline (This figure is
  available in color in the electronic version.)
  \label{fig:g043_map}}
\end{figure}

\begin{figure}[H]
  \begin{center}
    \includegraphics[scale=0.6]{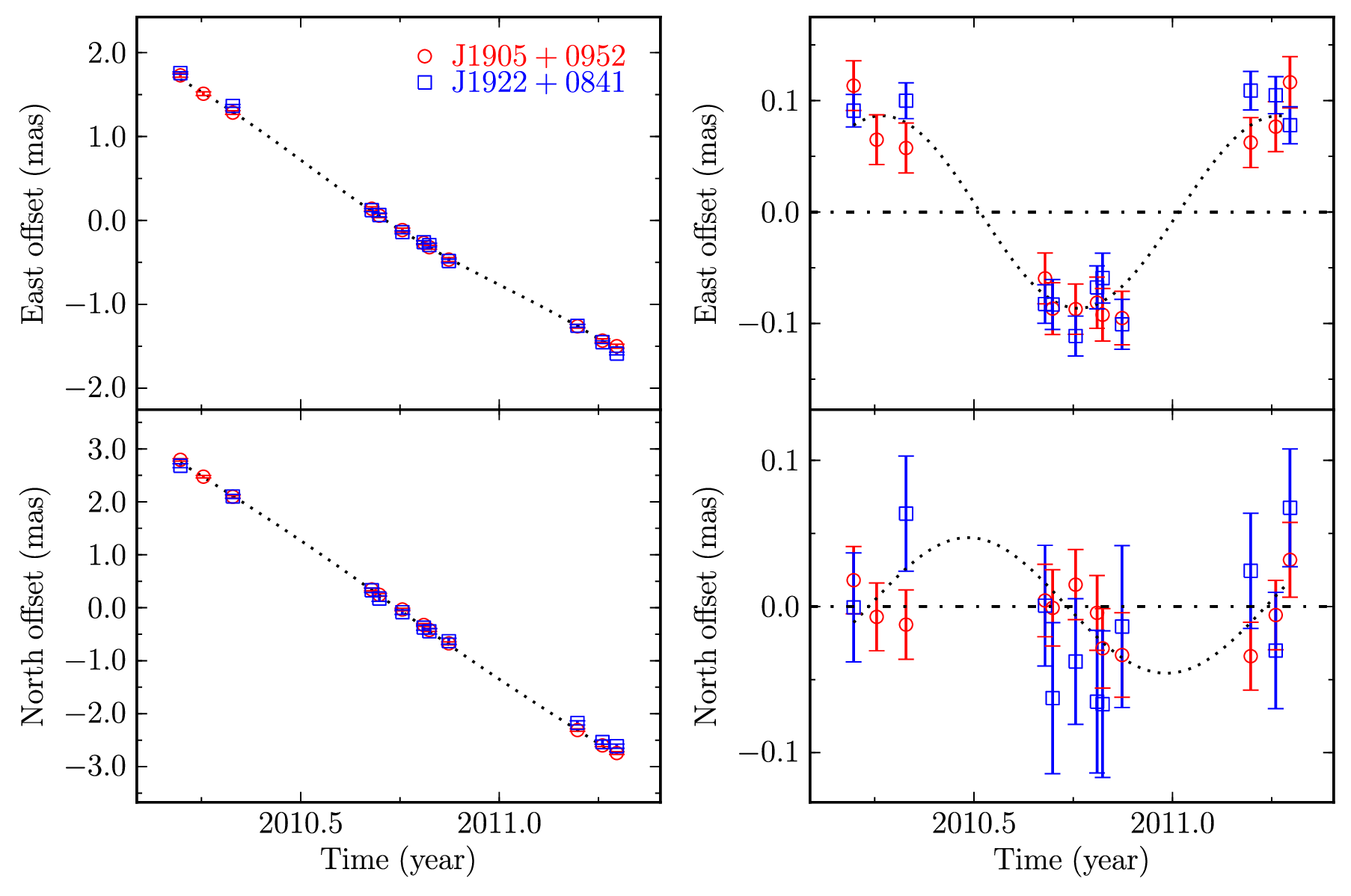}
  \end{center}
  \caption{
  Parallax and proper motion data ({\it markers}) and best-fitting model
  ({\it dotted line}) for the maser spot at the \VLSR\ of --44.77~\kms\ in
  W49N.  Plotted are positions of the maser spots relative to the
  extragalactic radio sources \Jfive\ ({\it circles}), \Jtwo\ ({\it
  squares}).  {\it Left panel}: Eastward ({\it upper panel}) and
  northward ({\it lower panel}) offsets versus time.  {\it Right panel}:
  Same as the {\it left panel}, except the best-fitting proper motion
  has been removed, displaying only the parallax signature.
  \label{fig:g043_para}}
\end{figure}

We found 11 maser spots detected at all twelve epochs that could be used
for precision astrometry.  These maser spots cluster in six locations
identified with letters A through F in Figure~\ref{fig:g043_map}.
Table~\ref{tab:g043_para_pm} shows the independent and combined parallax
fits for those maser spots.  Figure~\ref{fig:g043_para} shows the
independent parallax fit of the maser spot at \VLSR\ of --44.77 \kms\
with each background source as an example.  The combined parallax
estimate is $0.090 \pm 0.006$ mas, corresponding to a distance of
$11.11^{+0.79}_{-0.69}$~kpc.  The quoted parallax uncertainty is the
formal fitting error multiplied by $\sqrt{11}$, assuming conservatively
100\% correlated position uncertainties among the spots. 

Our distance to W49N is consistent with that of $11.4 \pm 1.2$ kpc
obtain by \citet{1992ApJ...393..149G} by modeling the expansion
(basically comparing maser Doppler velocities and proper motions).
Combining the data for the two background sources, we measured the
absolute proper motion of the reference maser spot to be \mux\ = --4.49
$\pm$ 0.13 \masy\ and \muy\ = --6.42 $\pm$ 0.12 \masy, where $\mu_x =
\mu_{\alpha}\cos\delta$ and $\mu_y = \mu_{\delta}$.

\begin{figure}[H]
  \begin{center}
    \includegraphics[scale=0.7]{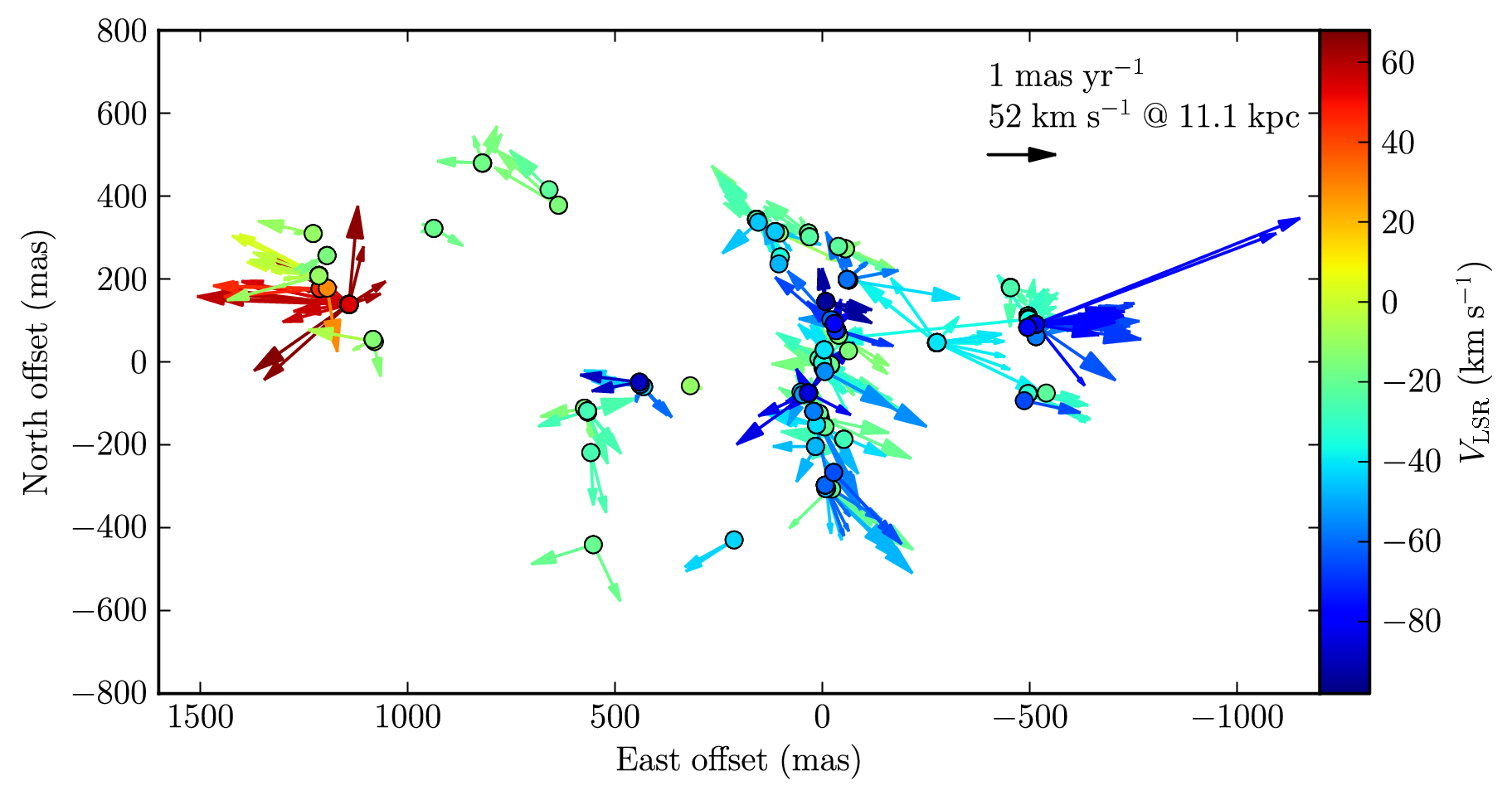}
  \end{center}
  \caption{
  Averaged positions ({\it circles}) and relative motions ({\it arrows})
  with mean value removed of maser spots respective to reference maser
  spot located at (0, 0) mas in W49N. The color bar denotes the \VLSR\
  range from --98 to 68 \kms\ of the maser features. The length and the
  direction of an arrow indicate the speed (given by the scale arrow in
  the upper right of the panel), the size of the arrow head is
  proportional to the uncertainty of the motion. \newline (This figure is
  available in color in the electronic version.)
  }
  \label{fig:g043_ipm}
\end{figure}

In order to model the internal motions of the masers to obtain the
motion of the exciting star(s), we used 345 maser spots which appeared
at four or more epochs within one year, and estimated their motions with
respective to the reference maser spot.  Figure~\ref{fig:g043_ipm} shows
the motions with their mean value removed, which indicates an expansion
originating from a small region that might include one (or more) young
stellar object(s), as suggested by \citet{1992ApJ...393..149G}.  We
fitted the data to models of expansion, with and without rotation.  The
estimated parameters from a Bayesian fitting procedure described by
\citet{2010ApJ...720.1055S} are listed in Table~\ref{tab:fit_g043}.

The motion of the expansion center relative to the reference maser spot
from the two models are in good agreement within their joint
uncertainties.  We adopt the results from the simpler model A as the
best estimate of the systematic motion.  Converting the ($V_{0x}$,
$V_{0y}$) to angular motions yields \mux\ = 2.01 $\pm$ 0.08 and \muy\ =
1.16 $\pm$ 0.06 \masy\ at the parallax distance of 11.1 kpc to W49N.
Adding this motion to the absolute motion of the reference maser spot,
we estimate an absolute proper motion of the exciting star(s) of W49N to
be \mux\ = --2.48 $\pm$ 0.15 \masy\ and \muy\ = --5.27 $\pm$ 0.13 \masy.
We note that the difference in proper motions of a maser spot
provided by different background sources are larger than their formal
errors for the individual fits.  This difference could result from small
structural variations such as jet motions in the background source or
from unmodeled atmospheric delays.

\subsection{\GVIII}

\begin{figure}[H]
  \centering
  \includegraphics[angle=0,scale=0.55]{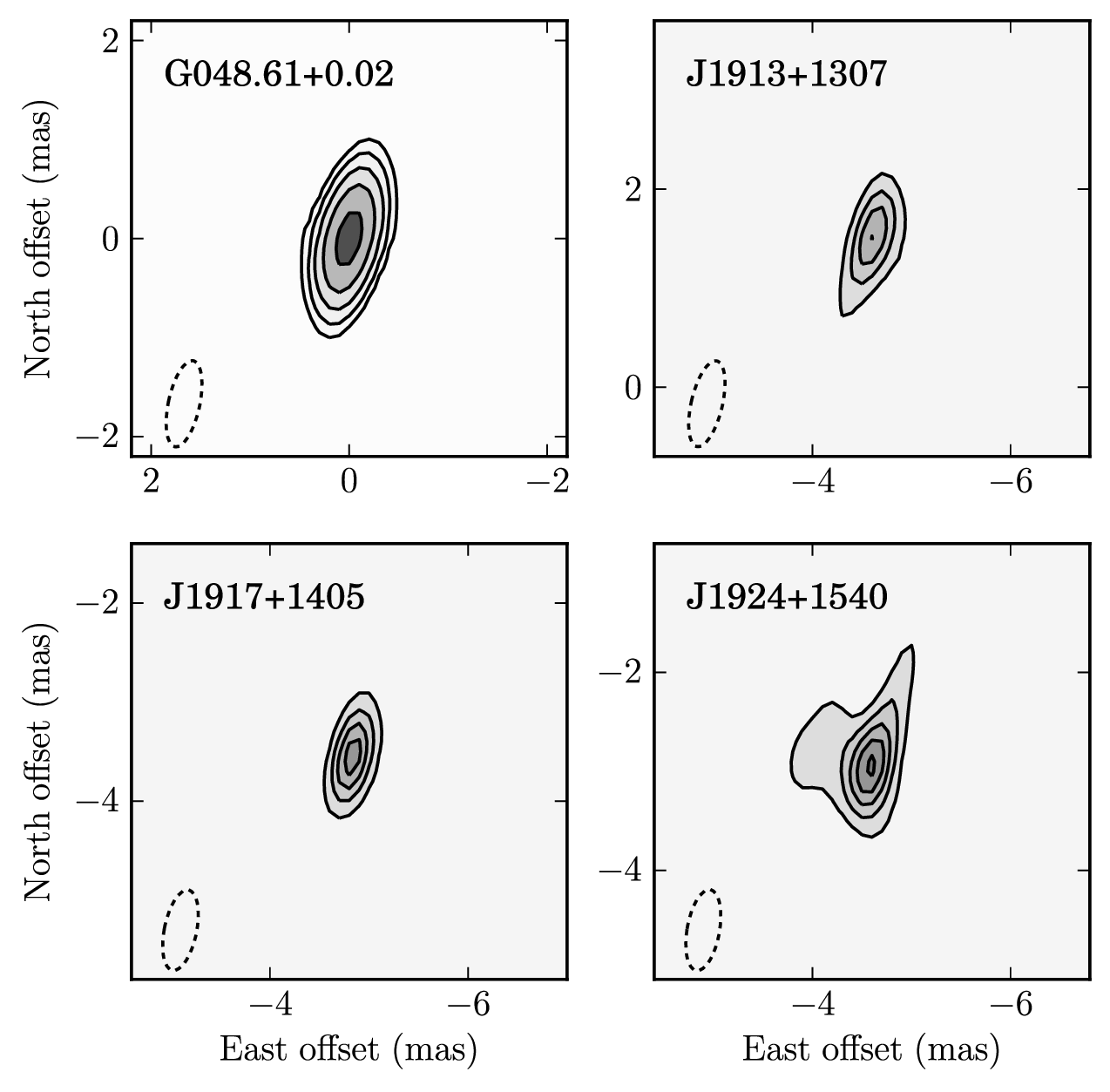}
  \caption{
  Images of the \hho\ maser reference spot at \VLSR\ of +25.58 \kms\ and
  the extragalactic radio sources used for the parallax measurements of
  \GVIII\ at the first epoch (2010 March 13).  Source names are in the
  upper left corner and the restoring beam (dotted ellipse) is in the
  lower left corner of each panel.  Contour levels for \GVIII\ maser
  emission are 5 \jybeam\ $\times~2^n$, $n$ = 0 \ldots 5, and for the
  background sources (\Jthree, \Jseven\ and \Jfour) are spaced linearly at
  0.01, 0.01 and 0.05 \jybeam, respectively.
  \label{fig:g048_mq}}
\end{figure}

\begin{figure}[H]
  \centering
  \includegraphics[angle=0,scale=0.6]{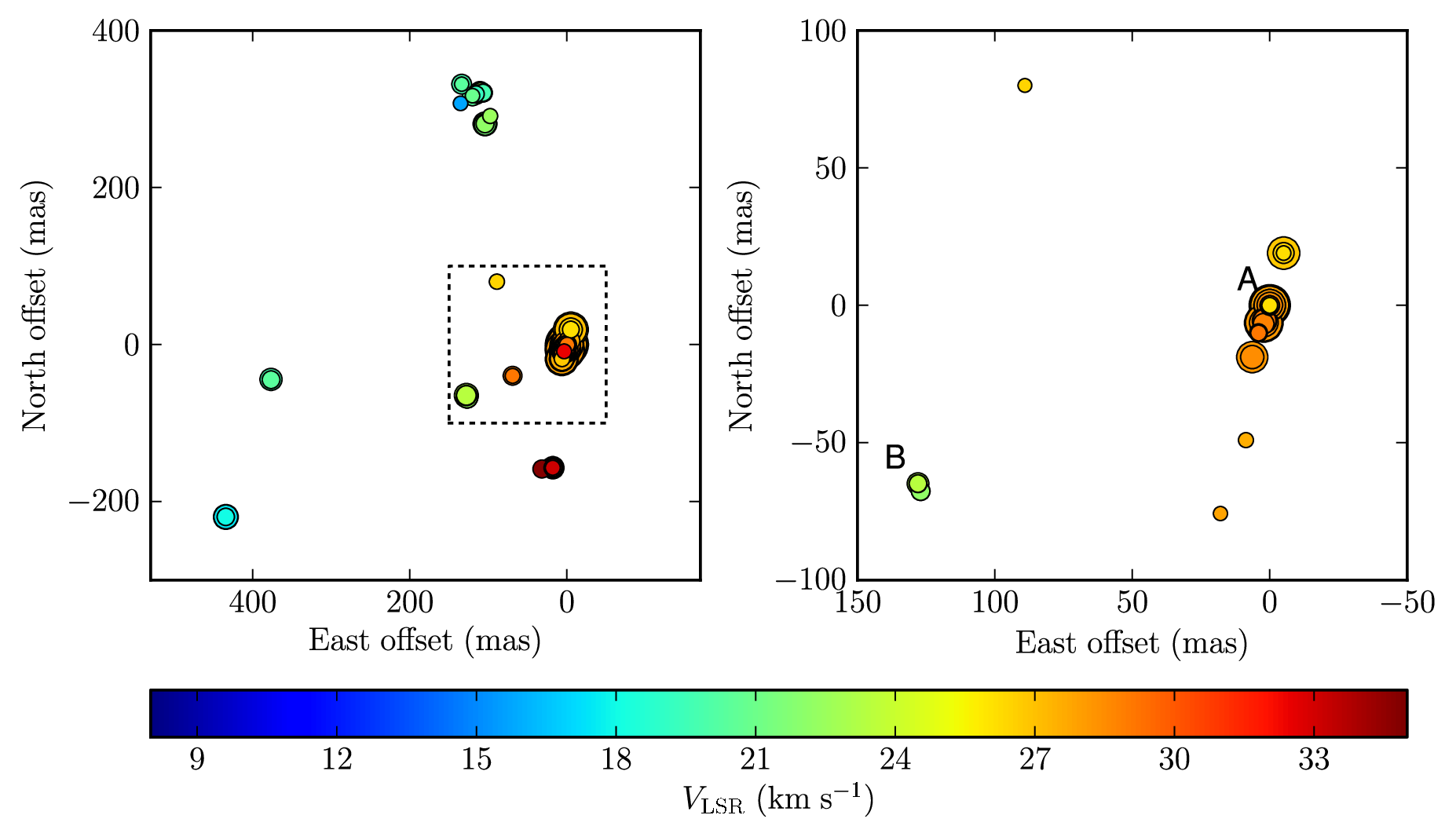}
  \caption{
  {\it Left panel}: spatial distribution of the \hho\ maser spots (with
  brightness $\ge$ 1 \jybeam\ for a restoring beam of $\sim 3 \times 1$
  mas) toward \GVIII\ from VLBA observations at the first epoch (2010
  March 13). Each maser spot is represented by a filled circle whose
  area is proportional to the logarithm of the flux density.  {\it Right
  panel}: blow-ups of the maser spot distribution in the field indicated
  with a dotted box in the {\it left panel}.  All the maser spots (with
  brightness $\ge$ 1 \jybeam\ for a restoring beam of $\sim 0.8 \times
  0.3$ mas) used for parallax fit are from this field.  Each region
  including maser spot for parallax fit is labeled with a letter.  The
  reference maser spot is in region A.  The color bar denotes the
  \VLSR\ range from +8 to +36 \kms. \newline  (This figure is available in color
  in the electronic version.)
  \label{fig:g048_map}}
\end{figure}

\begin{figure}[H]
  \begin{center}
    \includegraphics[scale=0.50]{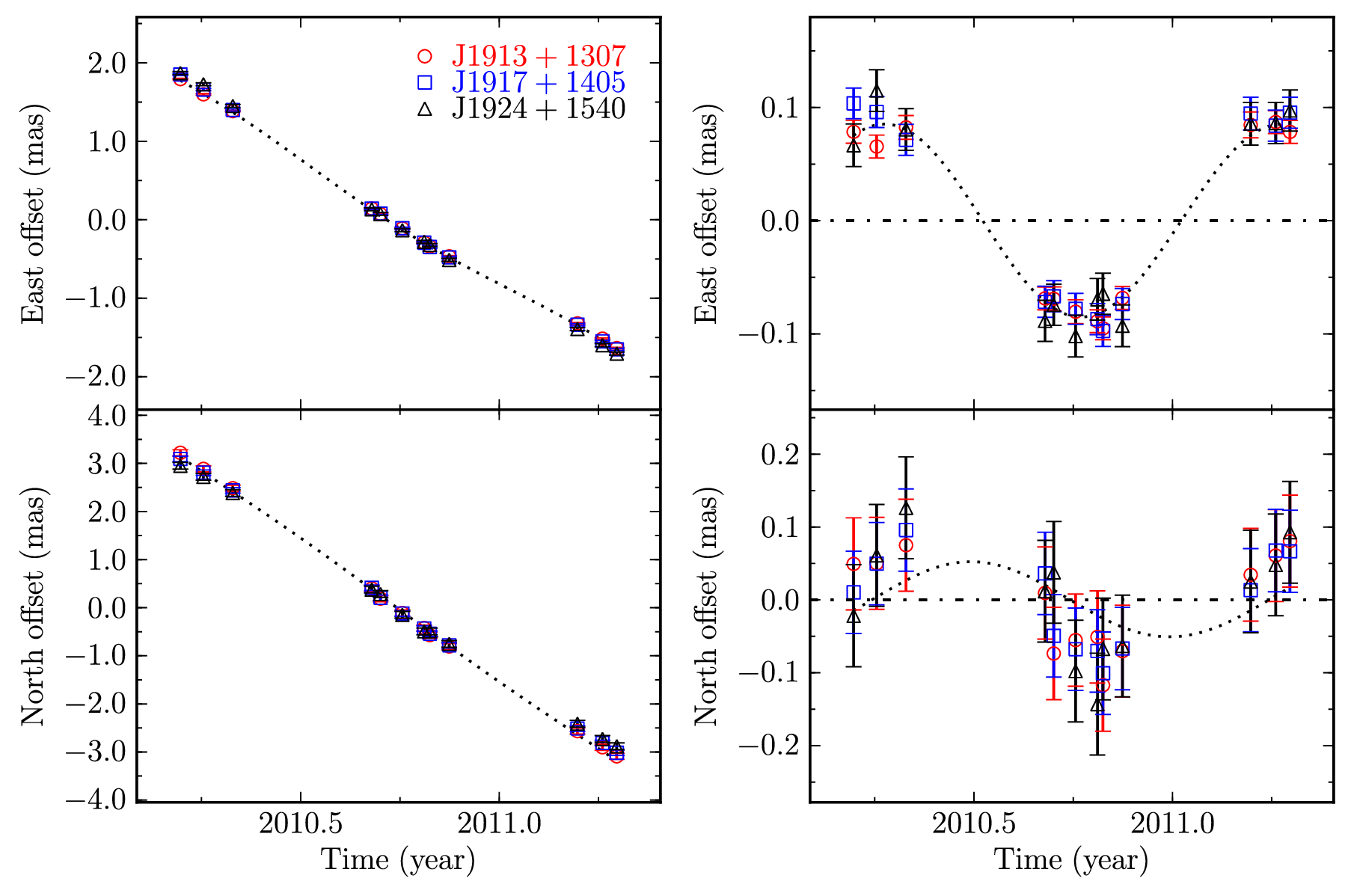}
  \end{center}
  \caption{
  Parallax and proper motion data ({\it markers}) and a best-fitting
  model ({\it dotted line}) for the maser spot at the \VLSR\ of
  +24.31~\kms\ in \GVIII.  Plotted are positions of the maser spot
  relative to the extragalactic radio sources \Jthree\ ({\it circles}),
  \Jseven\ ({\it squares}) and \Jfour\ ({\it triangles}).  {\it Left
  panel}: Eastward ({\it upper panel}) and northward ({\it lower panel})
  offsets versus time.  {\it Right panel}: Same as the {\it left panel},
  except the best-fitting proper motion has been removed, displaying
  only the parallax signature.
  \label{fig:g048_para}}
\end{figure}

For the parallax measurement of \GVIII, all the background sources were
detected at all epochs after phase-referencing to the maser spot at
\VLSR\ of +25.58 \kms.  Figure~\ref{fig:g048_mq} shows images of the
reference maser spot and the background sources at epoch 1.
Figure~\ref{fig:g048_map} shows the spatial distribution of \hho\ maser
spots and regions with maser spots used for the parallax fit.  We found
9 maser spots that appeared at all epochs, which could be used for
astrometric measurements.  Table~\ref{tab:g048_para_pm} shows the
independent and combined solution for those maser spots.
Figure~\ref{fig:g048_para} shows the parallax fit of
maser spot at \VLSR\ of +24.31 \kms\ with each background source as an
example.  The combined parallax estimate is $0.093 \pm 0.005$ mas,
corresponding to a distance of $10.75^{+0.61}_{-0.55}$~kpc. The quoted
parallax uncertainty is the formal fitting error multiplied by $\sqrt
{9}$, allowing for the possibility that position uncertainties of
maser spots are entirely correlated.  The absolute proper motion of the
reference maser spot is estimated to be \mux\ = --3.19 $\pm$ 0.01 \masy\
and \muy\ = --5.61 $\pm$ 0.03 \masy.

\begin{figure}[H]
  \begin{center}
    \includegraphics[scale=0.50]{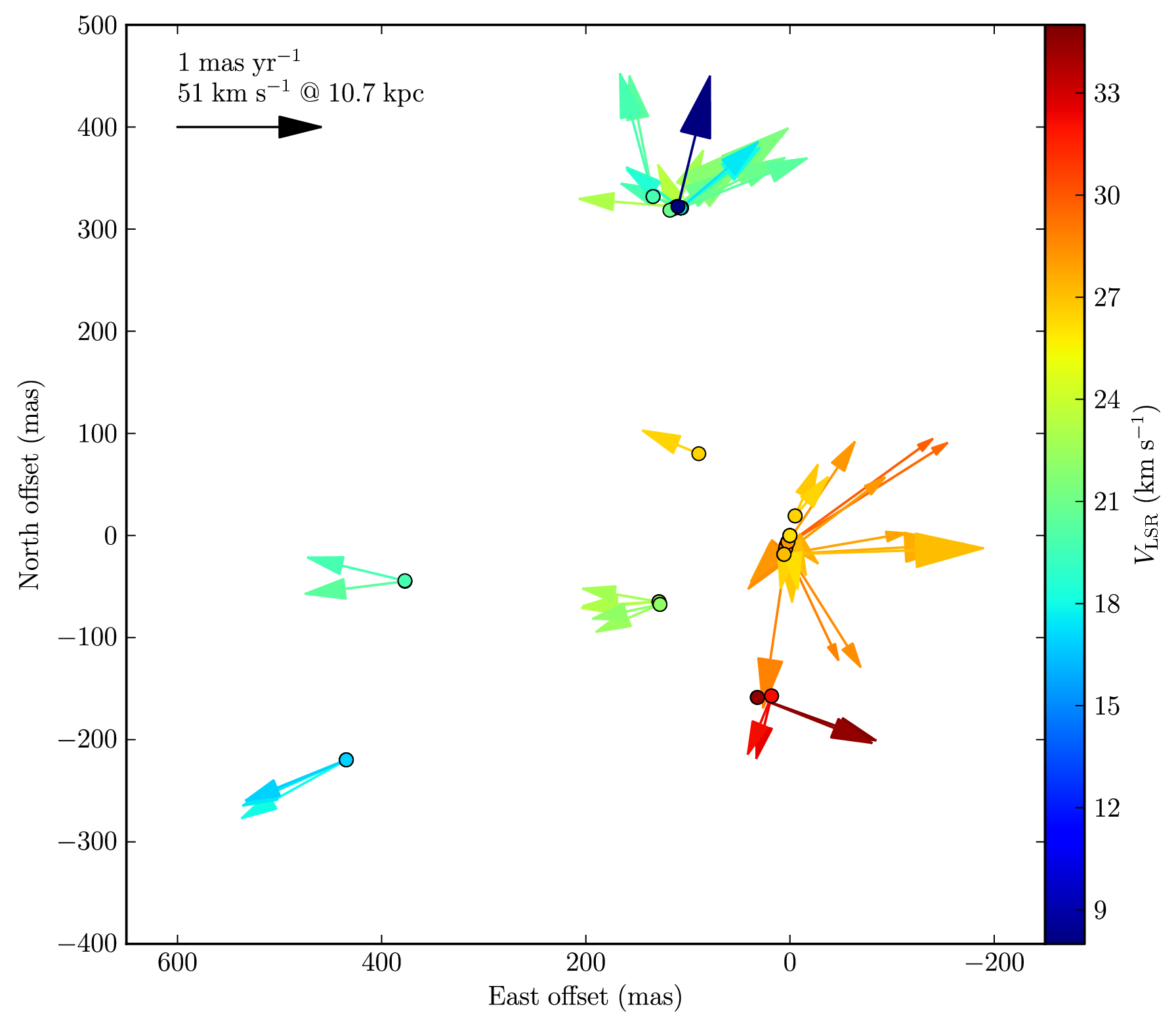}
  \end{center}
  \caption{ 
  Averaged positions ({\it circles}) and relative motions ({\it arrows})
  with mean value removed of maser spots relative to reference maser
  spot located at (0, 0) mas in \GVIII. The color bar denotes the \VLSR\ range
  from 8 to 35 \kms\ of the maser features. The length and the direction
  of an arrow indicate the speed (given by the scale arrow in the upper
  left of the panel), the size of the arrow head is proportional to the
  uncertainty of the motion. \newline (This figure is available in color in
  the electronic version.)
  }
  \label{fig:g048_ipm}
\end{figure}

We found 55 maser spots that were detected in at least four epochs over
one year and estimated their relative motions with respect to the
reference maser spot.  Figure~\ref{fig:g048_ipm} shows these motions
with their mean value removed. Similar to the kinematic modeling of
W49N, we fitted the data to expansion models with and without rotation.
The estimated parameters are listed in Table~\ref{tab:fit_g048}. Owing
to the large uncertainty of the parameters for rotation, we adopted the
results from model A.  The proper motion of the expansion center
relative to the reference maser spot corresponds to \mux\ = 0.30 $\pm$
0.08 \masy\ and \muy\ = 0.11 $\pm$ 0.08 \masy\ at a distance of 10.7
kpc.  Adding this to the absolute motion of the reference maser spot, we
obtained an absolute proper motion of the exciting star of \GVIII\ to be
\mux\ = --2.89 $\pm$ 0.08 \masy\ and \muy\ = --5.50 $\pm$ 0.09 \masy.

\section{DISCUSSION}
\label{sec:discuss}

\subsection{Inner portion of the Perseus arm traced by \hho\ maser sources}
\label{ssec:arm}

\begin{figure}[H]
  \begin{center}
    \includegraphics[scale=0.55]{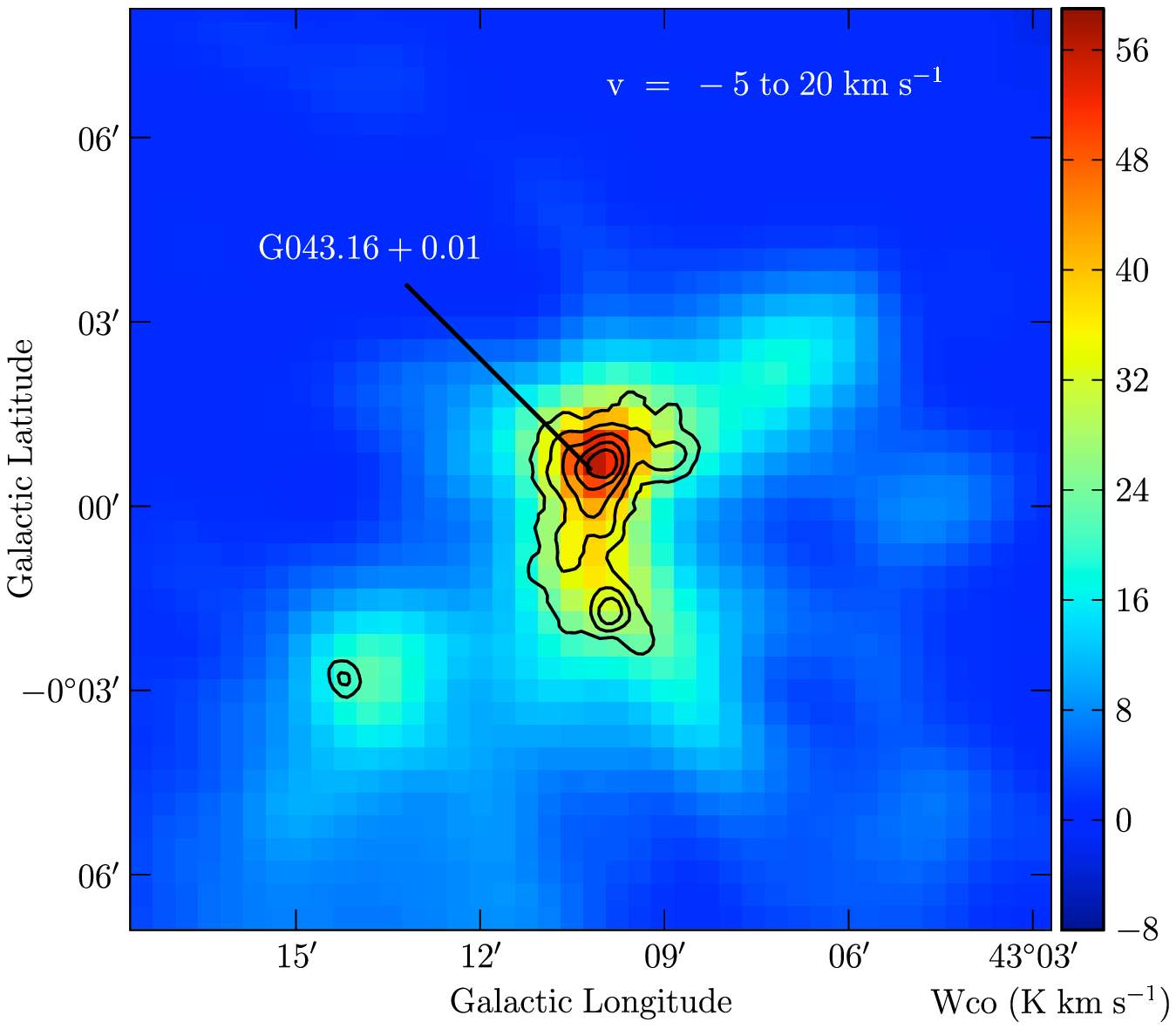}
    \hspace{3mm}
    \includegraphics[scale=0.55]{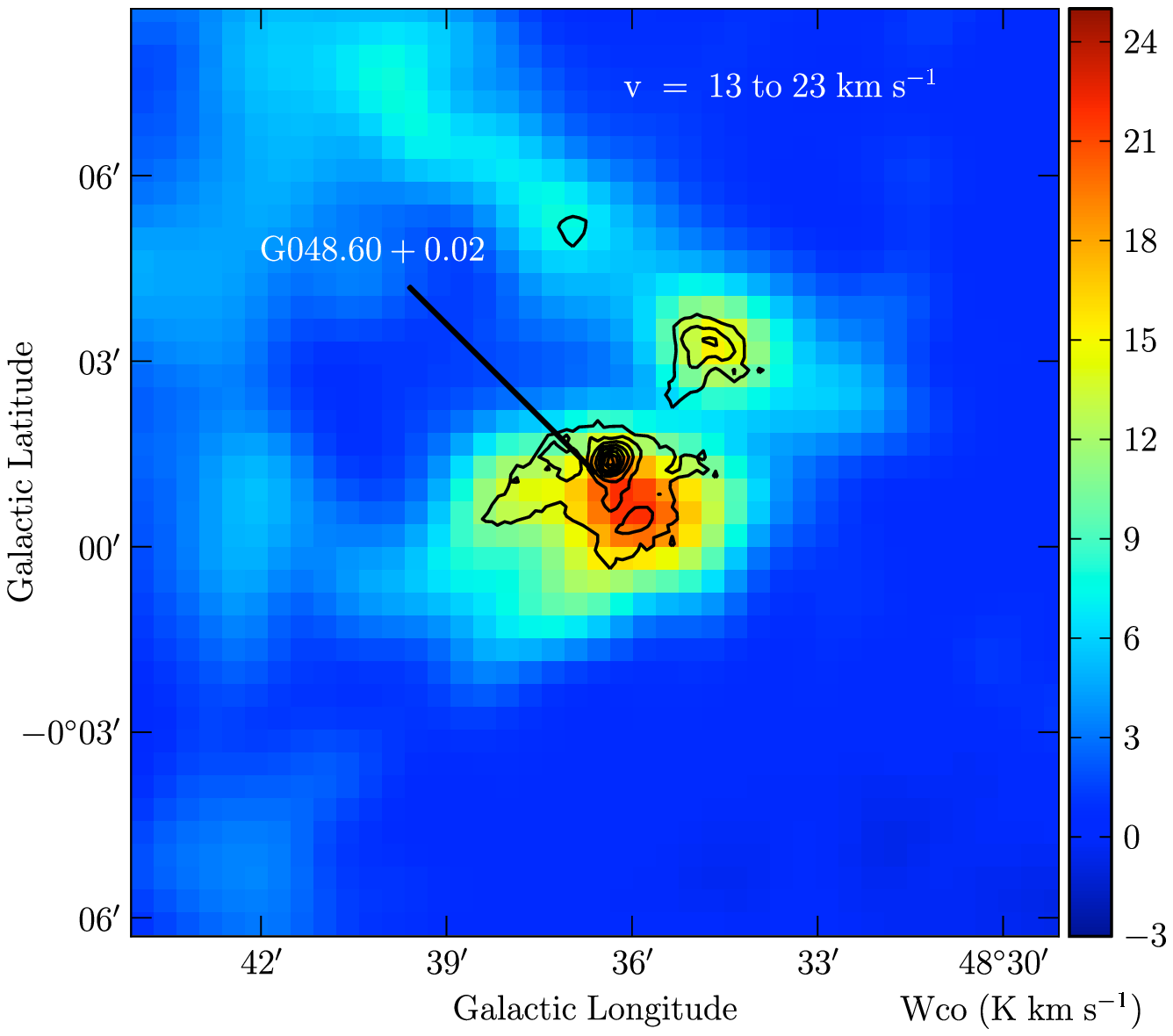}
  \end{center}
  \caption{
  Velocity-integrated \coIII\ intensity for the molecular clouds
  apparently associated \GIII\ ({\it left panel}) and \GVIII\
  ({\it right panel}). The velocity
  integration ranges are indicated in the upper right of each figure.
  The data are from the Galactic Ring
  Survey~\citep{2006ApJS..163..145J}. Over plotted contours are for the
  870 $\mu$m continuum emission from the APEX Telescope Large Area
  Survey of the Galaxy (ATLASGAL)~\citep{2009A&A...504..415S}. Contour
  levels are start at 2 \jybeam\ and increase by factors of 2 for \GIII;
  they are linearly spaced at 1 \jybeam\ for \GVIII.  \newline (This
  figure is available in color in the electronic version.)
  \label{fig:wco}}
\end{figure}

Spiral arms in the Galaxy can be identified as large-scale features in
longitude-velocity ($\ell-v$) diagrams from CO surveys (e.g.,
\citealt{1986ApJ...305..892D}).  We therefore assign our masers to arms
by associated them with molecular clouds, without possible bias by
using distance and a model of spiral structure.  A secure arm
assignment requires that the position and velocity of the maser and the
molecular cloud be in agreement with the position-velocity ($\ell-v$)
locus of the arm.  Using the data from the \coIII\ Galactic Ring Survey
(GRS) by \citet{2006ApJS..163..145J} and the APEX\footnote{APEX, the
Atacama Pathfinder EXperiments, is a collaboration between the Max
Planck Institut f\"ur Raodioastronomie, the Onsala Space Observatory,
and the European Southern Observatory.} Telescope Large Area Survey of
the GALaxy (ATLASGAL, \citealt{2009A&A...504..415S}), we determined that
both \GIII\ and \GVIII\ are nearly coincident on the sky with giant
molecular clouds (see Figure~\ref{fig:wco}).  The clouds have small
angular sizes ($\approx$ 3\arcmin), relatively large composite
linewidths (FWHM 9.4 and 5.0 \kms, respectively), and low positive
LSR velocities (11.1 and 17.7 \kms, respectively) which are consistent
with the $\ell-v$ locus identified by \citet{2008AJ....135.1301V} for
the Perseus arm.

\begin{figure}[H]
  \begin{center}
    \includegraphics[scale=0.55]{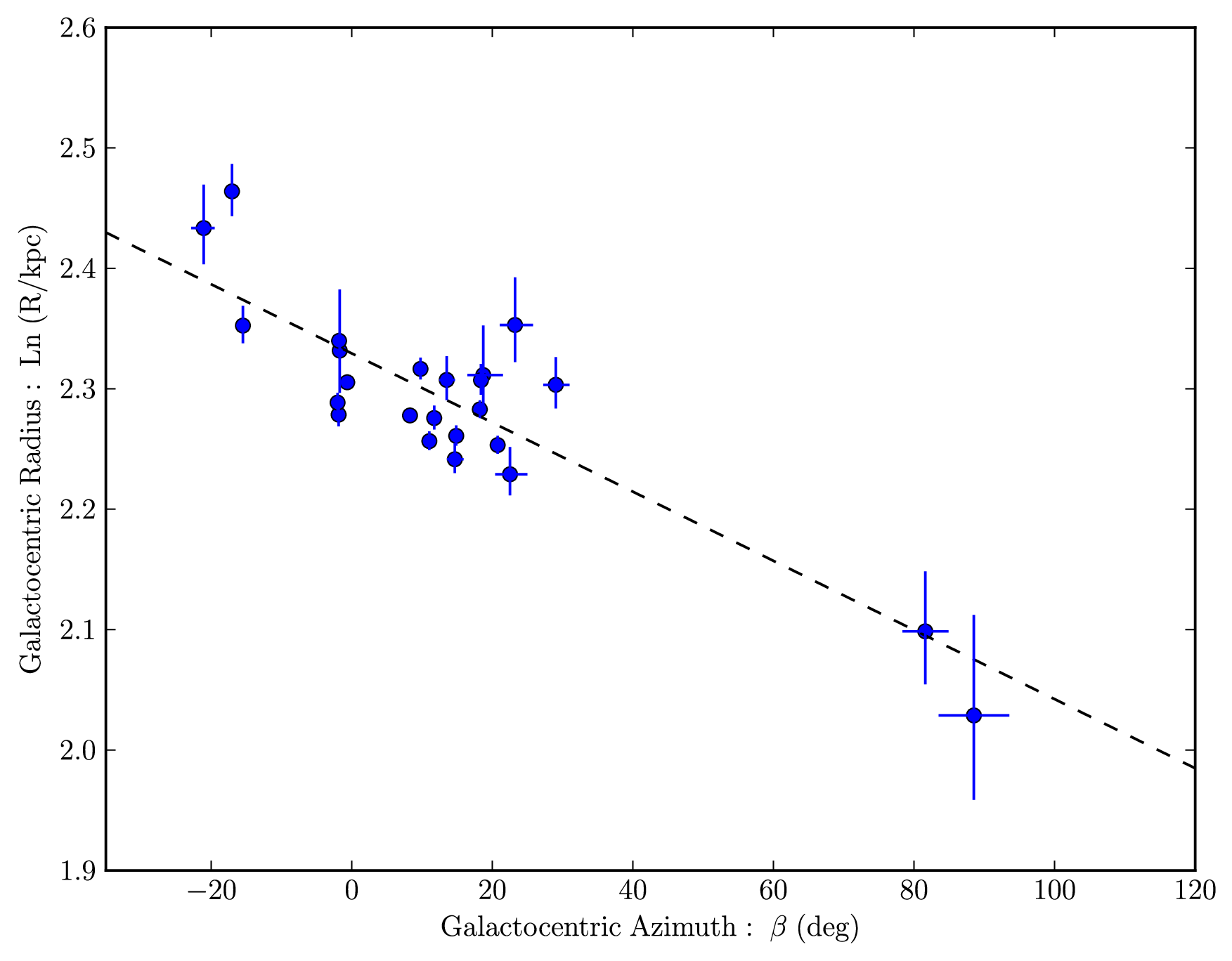}
  \end{center}
  \caption{
  Pitch angle of Perseus arm. The logarithm of Galactocentric radius,
  $R$ (in kpc units) is plotted against Galactocentric longitude
  ($\beta$).  The dashed line is a weighted fit to all the points and
  corresponds to a pitch angle of 9.5\deg.
  \label{fig:pitch_ang}}
\end{figure}

\begin{figure}[H]
  \begin{center}
    \includegraphics[scale=0.60]{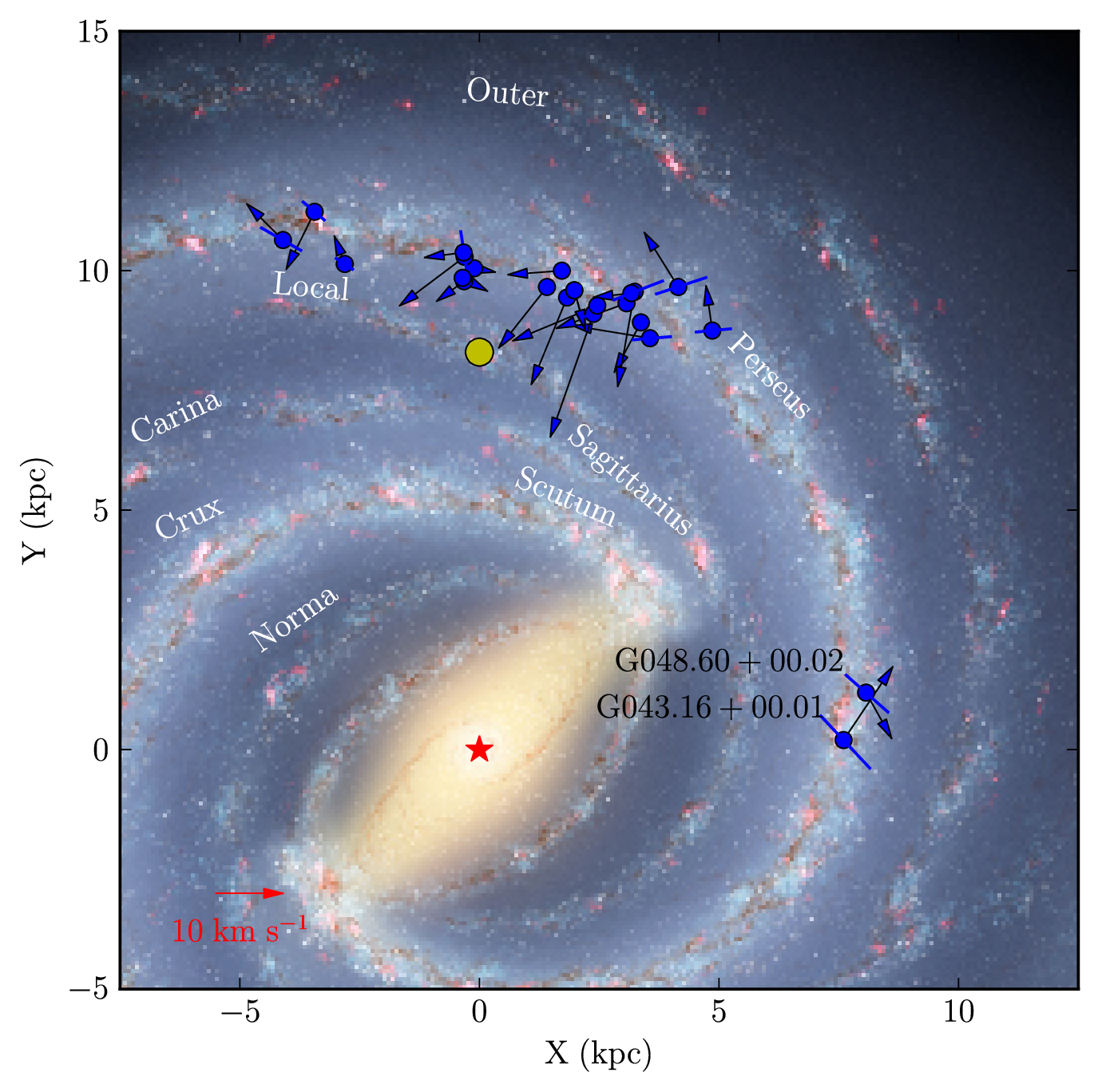}
  \end{center}
  \caption{
  Location ({\it dots with bar}) and peculiar motion ({\it arrows})
  projected on the Galactic plane  of \hho\ maser sources  with parallax
  measurements in the Perseus arm. The length of the bar denotes the
  distance uncertainty. A 10 \kms\ motion scale is in the lower left.
  The background is an artist's conception of Milky Way (R. Hurt:
  NASA/JPL-Caltech/SSC) viewed from the north Galactic pole from which
  the Galaxy rotates clockwise.  The Galactic center ({\it red star}) is
  at (0, 0) and the Sun ({\it yellow dot}) at (0, 8.3) kpc.  \newline
  (This figure is available in color in the electronic version.)
  \label{fig:mw_pos}}
\end{figure}

Combining the two sources in the inner portion of the Perseus arm
reported here with 22 sources with parallax measurements in the outer
portion of the Perseus arm from the BeSSeL Survey, the Japanese VLBI
Exploration of Radio Astrometry (VERA) project and the European VLBI
Network (Y.~K. Choi et al. 2013, in preparation), there are
now 24 sources defining the locus of the Perseus arm.
Figure~\ref{fig:pitch_ang} plots log($R$) versus $\beta$ for these
sources, where $R$ and $\beta$ are Galactocentric radius and azimuth
($\beta$ is defined as 0\deg\ toward the Sun and increasing with
Galactic longitude). As shown in Figure~\ref{fig:mw_pos}, these sources
are consistent with following a spiral from $\beta$ $\approx$ --25\deg\
to 90\deg\ (corresponding to Galactic longitude $\ell \approx$ 240\deg\
to 43\deg) and extending nearly 15 kpc in length.

Using a Bayesian fitting approach that takes into account uncertainties
in distance that maps into both R and $\beta$
(M.~J. Reid 2013 et al. 2013, in preparation)
and is insensitive to outliers \citep[see
``coservative formulation'']{2006OUP.book......S}, we estimate a global
pitch angle of 9.5\deg\ $\pm$ 1.3\deg\ for the Perseus arm, which is in
good agreement with that of 8.9\deg\ $\pm$ 2.1\deg\ determined from the
22 sources confined to the outer portion of the Perseus arm.  As we can
see from Figure~\ref{fig:pitch_ang}, \GIII\ and \GVIII\ at $\beta$ $\ge$
80\deg\ are crucial constraints for fitting a global pitch angle, since
most of the sources are located at $\beta$ $\le$ 25\deg. 
The preliminary pitch angle based on four sources in the Perseus
arm in \citet{2009ApJ...700..137R} of 16.5\deg\ $\pm$ 3.1\deg\ employed
a simpler least-squares fitting approach in which only the residual in
Galactocentric radius was minimized (when fitting a straight line to
log($R$) versus $\beta$).  Refitting the data from the four sources
available in \citet{2009ApJ...700..137R} with the new Bayesian approach,
which accounts for error in both log($R$) and $\beta$, yields 15.1\deg\
$\pm$ 6.1\deg.  Thus the preliminary fit and our new fit, based on 24
sources, agree within their joint uncertainty.

\subsection{3D motion in the inner portion of the Perseus arm}

Combining the parallax and proper motion measurements with the systemic
\VLSR\ (as listed in Table~\ref{tab:ppm}) enable us to determine the
three-dimensional (3D) peculiar motions (relative to circular motion
around the Galactic center) of \GIII\ and \GVIII. The \VLSR\ of \GIII\
estimated from maser kinematics modeling (see Table~\ref{tab:fit_g043})
is about 11 \kms, which is consistent with that of 11 \kms\ estimated
for the molecular cloud.  Similarly for \GVIII\ we estimate a \VLSR\ of
20 \kms\ from kinematic modeling of masers (see
Table~\ref{tab:fit_g048}), which is also close to the cloud velocity of
18 \kms.  To allow for a difference of \VLSR\ between \hho\ and CO,
we assign a \VLSR\ uncertainty of 5 \kms.  Similarly, there might be
additional uncertainty referring those maser motions to that of the
central star as reported in \S~\ref{sec:results}.  To account for this,
we added a proper motion error floor corresponding to 5 \kms\ to the
measurement uncertainty in quadrature. The final adopted astrometric
parameters and their uncertainties are listed in Table~\ref{tab:ppm}.

Assuming a flat rotation curve for the Milky Way with a rotation speed
of LSR \To\ = $239$ \kms, the distance to the Galactic center of
\Ro\ = $8.3$ kpc \citep{2011AN....332..461B}, and the solar motion of
(\U=$11.1$, \V=$12.24$, \W=$7.25$) \kms\ from revised Hipparcos
measurements by \citet{2010MNRAS.403.1829S}, we estimated peculiar
motions for the sources using the
procedure described in \citet{2009ApJ...700..137R}.  In order
to obtain realistic uncertainties for peculiar motions, we include the
effects of uncertainties in parallax, proper motion and \VLSR.  We do
this by generating 10,000 random trials consistent with our measured
values and (Gaussianly distributed) uncertainties.  The (\Us,\Vs,\Ws)
components of peculiar motion toward the Galactic center, in
the direction of Galactic rotation, and toward the north Galactic pole,
respectively, are given in Table~\ref{tab:ppm}.  All components are
consistent with zero motion, albeit with fairly large uncertainties
owing to the great distances of the sources.

\subsection{Distance to \GVIII}


%
%

\citet{2011PASJ...63..719N} also measured an \hho\ maser parallax
distance for \GVIII\ of $5.03 \pm 0.19$ kpc using the VERA array.  This
result is significantly different from our distance of
10.75$^{+0.61}_{-0.55}$ kpc.  The Nagayama distance would place
\GVIII\ in the Sagittarius-Carina arm and near (projected separation
$\approx$ 1\deg) the supernova remnant G049.49-0.37 and the active
star-forming region W51, which has several parallax distance
measurements near
5.3~kpc~\citep{2009ApJ...693..413X,2010ApJ...720.1055S,2013Submitted.....W}.
Note that the \VLSR\ of 18 \kms\ for \GVIII\ is offset by
$\approx50$ \kms\ from the W51 sources.  This is inconsistent with the
spiral arm assignment of \GVIII\ to the inner portion of the Perseus arm
(described in \S~\ref{ssec:arm} and based on \coIII\ position--velocity
information).  \citet{2011PASJ...63..719N} suggested that the large
velocity offset might be the result of the multiple SN explosions in
W51.  However, this is inconsistent with N-body simulations by
\citet{2009ApJ...706..471B}, which suggest that the acceleration by the
SN explosion does not cause motions of that magnitude for swept-up,
star-forming gas.

\begin{figure}[H]
  \begin{center}
    \includegraphics[scale=0.5]{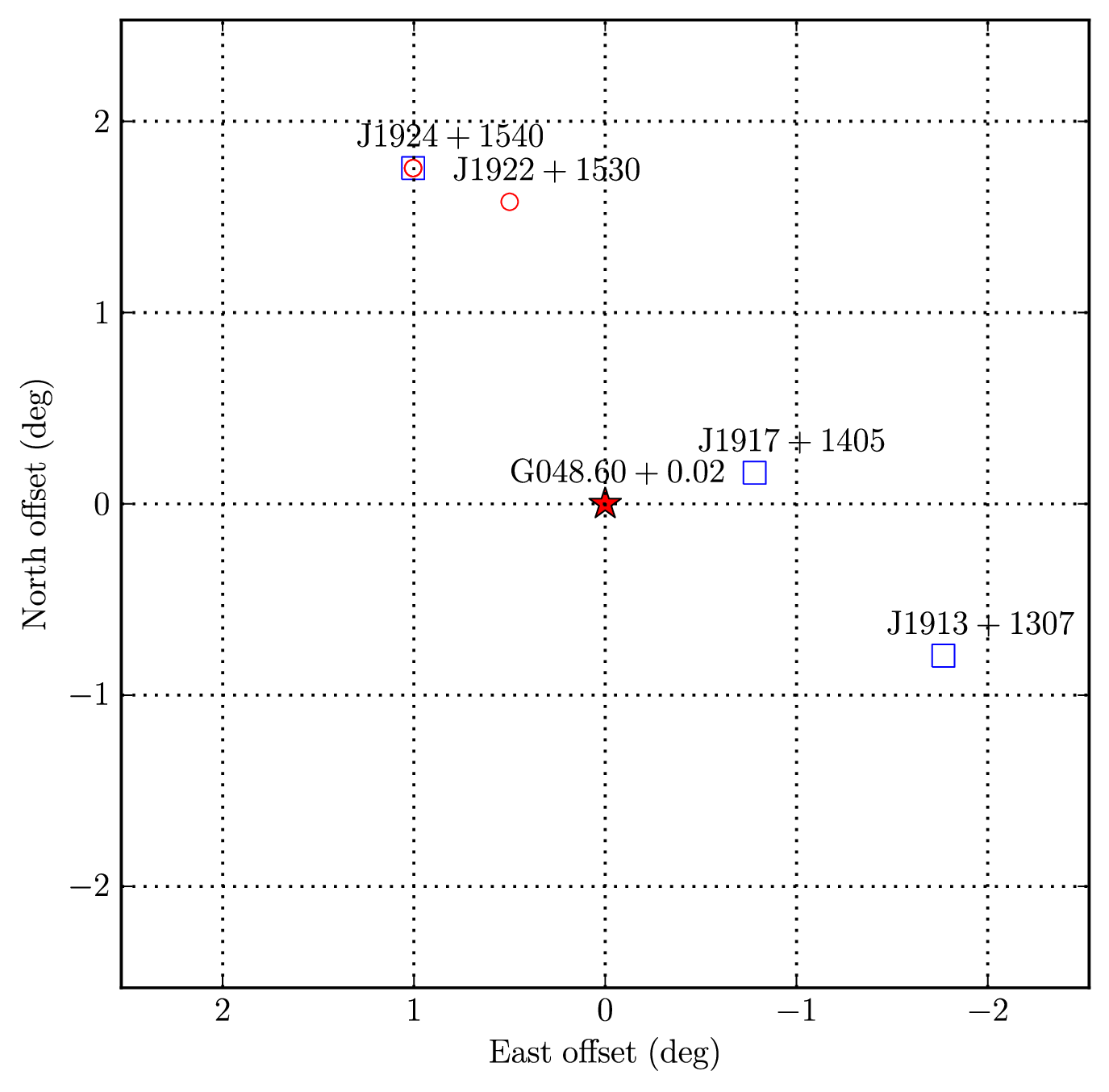}
  \end{center}
  \caption{
  Sky positions of the target and background sources for \GVIII. {\it
  Squares/Circles} denote background sources used in the VLBA/VERA
  observations.
  }
  \label{fig:srcpos}
\end{figure}

Generally there is excellent agreement between parallaxes measured by
different VLBI arrays.  What could explain this unusual difference?  Our
VLBA parallax measurement has some superiorities over those of
\citet{2011PASJ...63..719N}.  Firstly, our observations have longer
interferometric baselines.  Secondly, our observations have more and
closer background sources, nearly symmetrically distributed relative to
the target as shown in Figure~\ref{fig:srcpos}.  This could be very
important to reduce systematic error due to unmodeled tropospheric
delays.  Note that the background sources used by Nagayama are very
close together and both are offset mostly north of \GVIII.   Therefore,
unmodeled tropospheric delays will be similar for both of their
background sources.  Noting this source of correlation, as well as
the nearly 100\% correlation among different maser spot positions,
owing to nearly identical unmodeled tropospheric delays, the (formal)
parallax uncertainty quoted by Nagayama of $\pm0.007$ mas, may be
underestimated by factors of $\sqrt{9}$ and $\sqrt{2}$ (for 9 maser
spots and 2 background sources) and more realistically is $\pm0.030$
mas.  However, even with this uncertainty, the difference between our
and Nagayama's parallaxes ($0.103\pm0.031$) is still somewhat
significant.  Thirdly, our observing epochs symmetrically sample the
peaks of the sinusoidal parallax signature in right ascension, yielding
near-zero correlation coefficients between parallax and proper motion
terms.  For these reasons, we suspect that our measured distance
of 10.75$^{+0.61}_{-0.55}$ kpc to \GVIII\ is more reliable.

\subsection{A Star Formation Gap in the Perseus arm}

\begin{figure}[H]
  \begin{center}
    \includegraphics[scale=0.75]{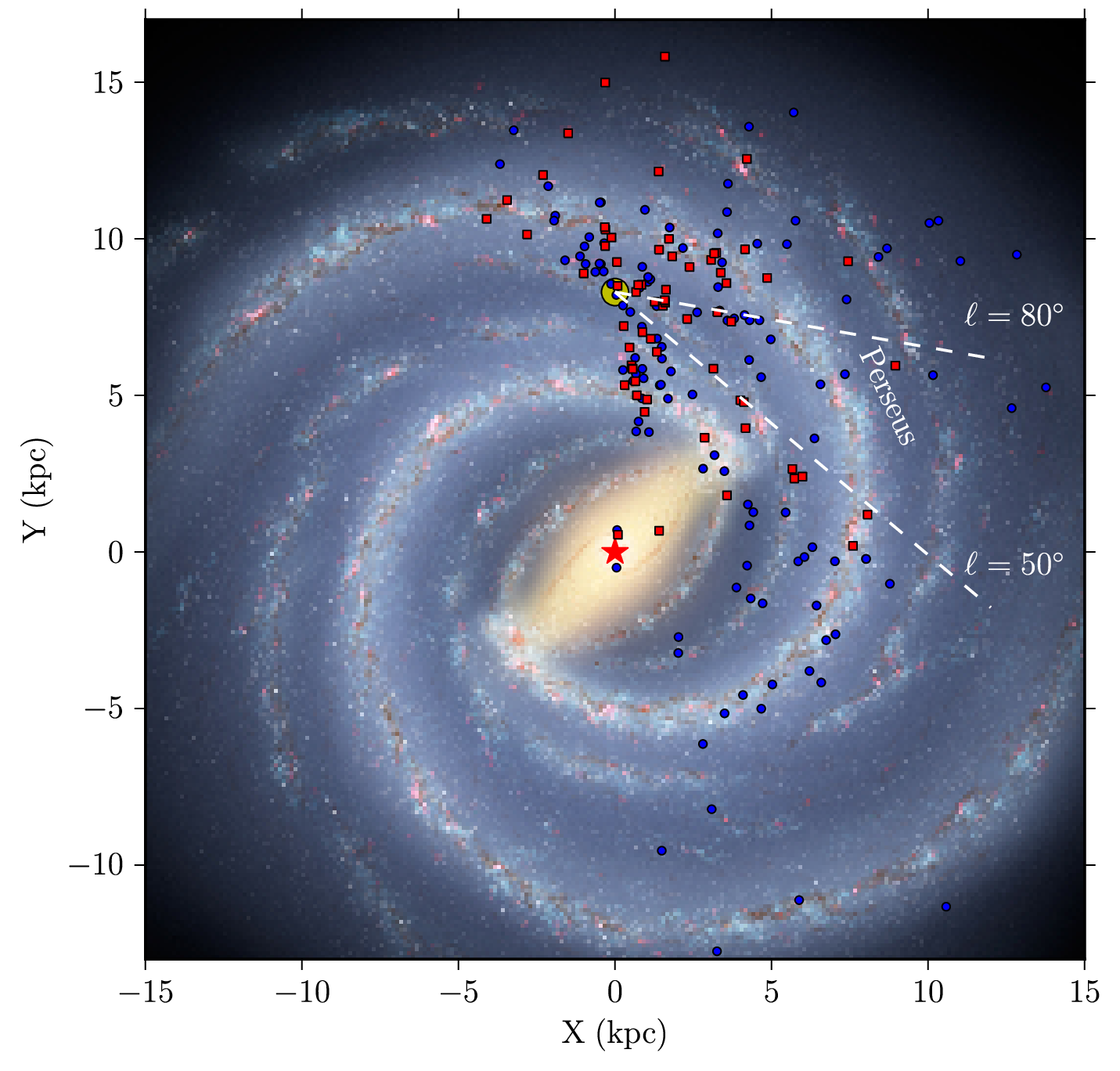}
  \end{center}
  \caption{
  Galactic distributions of 22 GHz \hho\ masers with kinematic distances
  ({\it blue dots}) from \citet{2001A&A...368..845V} and parallax
  distances ({\it red squares}) from M.~J.~Reid et al. (2013, in preparation).
  Kinematic distances have been used to generate these plots and
  distance ambiguities have been resolved using the prescription of
  \citet{2003ApJ...587..701F}, based simply on Galactic latitude. Note
  that these kinematic distances are highly uncertain and are almost
  useless to determine spiral structure.  The background is an artist's
  conception of Milky Way (R. Hurt: NASA/JPL-Caltech/SSC) viewed from
  the north Galactic pole from which the Galaxy rotates clockwise.  The
  Galactic center ({\it red star}) is at (0, 0) and the Sun ({\it yellow
  dot}) at (0, 8.3) kpc. \newline (This figure is available in color in
  the electronic version.)
  \label{fig:mw_maser}}
\end{figure}

As shown in Figure~\ref{fig:pitch_ang}, we have yet to locate a
high-mass star forming region with $\beta$ between $\approx$ 30\deg\ and
80\deg\ in the Perseus arm.   Although we observed sources whose sky
positions and velocities suggested kinematic distances in
the Perseus arm, all were found to be much closer and located in the
Local arm~\citep{2013ApJ...769...15X}.  In Figure~\ref{fig:mw_maser}, we
plot the Galactic plane locations (based on kinematic distances) of
$\approx$ 200 22-GHz \hho\ masers stronger than 10 Jy and associated
with star-forming regions.  We find that the section of the Perseus arm
between $\ell \approx$ 50\deg\ to 80\deg\ near the Solar Circle has very
few 22 GHz candidate \hho\ masers, even though they are numerous
masers outside the Solar Circle in the Perseus arm for $\ell \ge$
90\deg.

\begin{figure}[H]
  \begin{center}
    \includegraphics[width=0.95\textwidth]{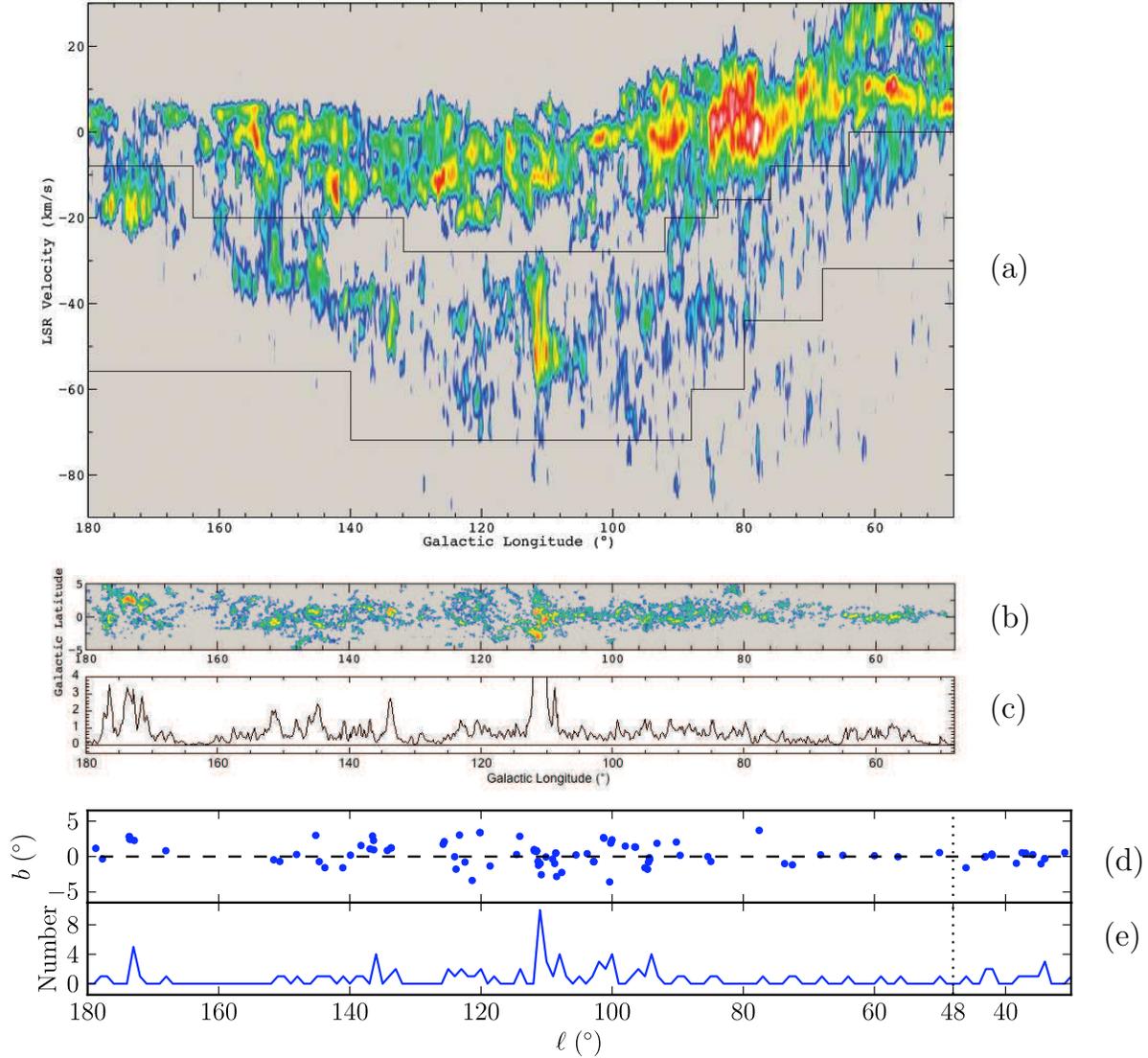}
  \end{center}
  \caption{
  (a): A CO longitude-velocity diagram integrated over a 10\deg\ strip
  of Galactic latitude centered on the Galactic plane from
  \citet{2001ApJ...547..792D}. The solid lines roughly outline the locus
  of the Perseus arm. The colors indicate log intensity, from 0.1
  K-arcdeg ({\it blue}) to 16 K-arcdeg ({\it white}). Below the right
  edge at longitude of 48\deg\ the Perseus arm is inside the solar
  circle with small positive velocity and blends with stronger, more
  extensive local emission, and hence it is not shown.
  (b): A velocity-integrated CO map of the Perseus arm, obtained by
  integrating the \citet{2001ApJ...547..792D} survey over the velocity
  ranges indicated in (a). The colors indicate log Wco, from 1 K \kms\
  (blue) to 100 K \kms\ (white).
  (c): Mean CO intensity of the Perseus arm vs. Galactic longitude, obtained by
  averaging the map in (b) over latitude.
  (d): Galactic longitude and latitude of Massive Young Stellar Objects
  (MYSOs) within the velocity ranges indicated in (a). These MYSOs are
  from the Red MSX Sources (RMS)
  Survey~(\url{http://www.ast.leeds.ac.uk/RMS/}).
  (e): Distribution of Galactic longitude of MSYOs plotted in (d). The
  dotted line in (d) and (e) indicates longitude of 48\deg.
  \newline (This figure is available in color in the electronic version.)
  \label{fig:per_lv}}
\end{figure}

As shown in Figure~\ref{fig:per_lv}, CO emission indicates that the
Perseus arm can be seen from $\ell \approx$ 180\deg to 48\deg, where it
spirals to the Solar Circle and \VLSR\ values approach zero and merge
with local emission.  The longitude range 64\deg\ to 76\deg\ is weak in
CO emission (as it is from 160\deg\ to 168\deg).  This suggests that the
Perseus arm, at least between $\ell \approx$ 64\deg\ and 76\deg, is low
in giant molecular clouds and massive star formation. Also plotted is
the Galactic distribution of Massive Young Stellar Objects (MYSOs) from
the Red MSX Sources (RMS) Survey \citep{2007A&A...461...11U}.  Since
MYSOs are indicators of star formation, the small numbers of MYSOs at
50\deg\ $< \ell <$ 80\deg\ also indicates low levels of star formation
in this portion of the Perseus arm.

\section{SUMMARY}

We measured trigonometric parallaxes and proper motions of \hho\ masers
in two star-forming regions, \GIII\ and \GVIII.   We establish that both
sources are at great distances and that \GIII\ is one of the most
luminous star forming regions in the Milky Way.  These two sources have
positions and \VLSR's that match \coIII\ emission from giant molecular
clouds that are located in the Perseus arm.  Thus, our parallax
distances accurately locate the inner portion of the Perseus arm within
the Milky Way.  Combining our results with other parallax measurements
for  maser sources associated with the outer portion of the Perseus arm,
we determined a global pitch angle of 9.3\deg\ $\pm$ 1.3\deg\ for the
Perseus arm.  Finally, we suggest that there is little massive-star
formation in the Perseus arm between $l \approx 50$\deg\ and 80\deg.

\begin{acknowledgements}

The work was supported by the National Science Foundation of China
(under grants 10921063, 11073046, 11073054 and 11133008) and the Key
Laboratory for Radio Astronomy, Chinese Academy of Sciences.  This work
was partially funded by the ERC Advanced Investigator Grant GLOSTAR
(247078).  We are grateful to Dr. James Urquhart for providing the
ATLASGAL FITS files.  We also acknowledge Dr. John D. Hunter, the
creator of the Python {\it matplotlib} which was used extensively in our
figures.

\end{acknowledgements}

{\it Facilities:} \facility{VLBA}



\onecolumn

\clearpage

\begin{deluxetable}{clclcrlc}
\tablecolumns{7} \tablewidth{0pc} \tablecaption{VLBA Observations}
\tablehead {
\colhead{Epoch} & \colhead{Program} & \colhead{Date} &  \colhead{Antennas Available} 
\\
\colhead{group} &  \colhead{code}   & \colhead{(yyyy mm dd)} &  \colhead{BR FD HN KP LA MK NL OV PT SC}
            }
\startdata
1 & BR145B1  & 2010 03 13  & \c~~~\c~~~\c~~~\c~~~\c~~~\c~~~\c~~~\c~~~\c~~~\c \\
1 & BR145B2  & 2010 04 03  & \c~~~\c~~~\c~~~\c~~~\c~~~\c~~~\c~~~\c~~~\c~~~\c \\
1 & BR145B3  & 2010 04 30  & \c~~~\c~~~\c~~~\c~~~\x~~~\c~~~\c~~~\c~~~\c~~~\c \\
2 & BR145B4  & 2010 09 05  & \c~~~\c~~~\c~~~\c~~~\c~~~\c~~~\c~~~\c~~~\c~~~\c \\
2 & BR145B5  & 2010 09 12  & \c~~~\c~~~\c~~~\c~~~\c~~~\c~~~\c~~~\c~~~\c~~~\c \\
2 & BR145B6  & 2010 10 03  & \c~~~\c~~~\c~~~\c~~~\c~~~\c~~~\c~~~\c~~~\c~~~\c \\
2 & BR145B7  & 2010 10 23  & \c~~~\c~~~\c~~~\c~~~\c~~~\c~~~\c~~~\c~~~\c~~~\c \\
2 & BR145B8  & 2010 10 28  & \c~~~\c~~~\c~~~\c~~~\c~~~\c~~~\c~~~\c~~~\c~~~\c \\
2 & BR145B9  & 2010 11 15  & \c~~~\c~~~\c~~~\c~~~\c~~~\c~~~\c~~~\c~~~\c~~~\c \\
3 & BR145BA  & 2011 03 13  & \c~~~\c~~~\c~~~\c~~~\c~~~\x~~~\c~~~\c~~~\c~~~\c \\
3 & BR145BB  & 2011 04 05  & \c~~~\c~~~\c~~~\x~~~\c~~~\c~~~\c~~~\c~~~\c~~~\c \\
3 & BR145BC  & 2011 04 18  & \c~~~\c~~~\c~~~\c~~~\c~~~\c~~~\x~~~\c~~~\x~~~\c \\
\enddata
\tablecomments {The first column lists the epoch group number, which
denotes the order number of peaks in the two year sinusoidal
trigonometric parallax signature. Check marks indicate that the antenna
produced good data, while a blank indicates that little or no useful
data was obtained.  Antenna codes are BR: Brewster, WA; FD: Fort Davis,
TX; HN: Hancock, NH; KP: Kitt Peak, AZ; LA: Los Alamos, NM; MK: Mauna
Kea, HI; NL: North Liberty, IA; OV: Owens Valley, CA; PT: Pie Town, NM;
and SC: Saint Croix, VI. }

\label{tab:obs}
\end{deluxetable}

\begin{deluxetable}{ccccrccc}
  \tablecolumns{8}
  \tablewidth{0pc}
  \tablecaption{Source characteristics}
  \tablehead{
  \colhead{Source} & \colhead{R.A. (J2000)} & \colhead{Dec. (J2000)}                & \colhead{$\theta_{sep}$} & \colhead{P.A.}    & \colhead{$S$}       & \colhead{\VLSR}  & \colhead{Beam}\\
  \colhead{     }  & \colhead{(h~~~m~~~s)}  & \colhead{(\degr~~~\arcmin~~~\arcsec)} & \colhead{(\degr)}        & \colhead{(\degr)} & \colhead{(\jybeam)} & \colhead{(\kms)} & \colhead{(mas~~mas~~\degr)}
  }
  \startdata
       \GIII & 19 10 13.4096 & +09 06 12.803 & ... &   ... &   1000\to8000 & --4.75 & 0.8 $\times$  0.4 @ \p--0 \\
  J1905+0952 & 19 05 39.8989 & +09 52 08.407 & 1.4 &  --56 & 0.050\to0.170 &        & 0.7 $\times$  0.3 @ \p--8 \\
  J1922+0841 & 19 22 18.6337 & +08 41 57.373 & 3.0 &   +98 & 0.010\to0.020 &        & 0.9 $\times$  0.3 @ \p--9 \\
             &               &               &     &       &               &        &                           \\
      \GVIII & 19 20 31.1761 & +13 55 25.209 & ... &   ... &      90\to160 & +25.58 & 0.8 $\times$  0.3 @ \p--9 \\
  J1917+1405 & 19 17 18.0641 & +14 05 09.769 & 0.8 &  --78 & 0.020\to0.050 &        & 0.8 $\times$  0.3 @  --13 \\
  J1913+1307 & 19 13 18.0641 & +13 07 47.331 & 1.9 & --114 & 0.010\to0.050 &        & 0.8 $\times$  0.3 @  --12 \\
  J1924+1540 & 19 24 39.4559 & +15 40 43.941 & 2.0 &   +30 & 0.090\to0.450 &        & 0.8 $\times$  0.3 @  --12 \\
\enddata
\tablecomments{
Column 1 gives the names of the maser sources and its corresponding
background sources. Columns 2 to 3 list the absolute positions of the
reference maser spot and background sources.  Columns 4 to 5 give the
separations ($\theta_{sep})$ and position angles (P.A.) east of north
between maser and background sources.  Columns 6 to 7 give the
brightnesses ($S$) and \VLSR\ of reference maser spot.  The last column
gives the full width at half maximum (FWHM) size and P.A. of the
Gaussian restoring beam.  Calibrators are from the BeSSeL calibrator
survey~\citep{2011ApJS..194...25I}.
}
\label{tab:src}
\end{deluxetable}

\setlength{\tabcolsep}{2pt}
\begin{deluxetable}{lccrrrrrr rrr }
 \tabletypesize{\footnotesize} 
\tablecaption{Parallax and proper motion fits for W49N~\label{tab:g043_para_pm}}
\tablewidth{0pt} 
\tablehead{
\colhead{Background}&\colhead{Region}&\colhead{\VLSR} &\colhead{Parallax}&\colhead{\mux}   &\colhead{\muy}   &\colhead{\dx}  &\colhead{\dy}  &\colhead{$\chi^2_\nu$}&\colhead{$\sigma_x$}&\colhead{$\sigma_y$} \\
    \colhead{source}&\colhead{}      &\colhead{(\kms)}&\colhead{(mas)}   &\colhead{(\masy)}&\colhead{(\masy)}&\colhead{(mas)}&\colhead{(mas)}&\colhead{}            &\colhead{(mas)} &\colhead{(mas)} }

\startdata

   \Jfive &   B &  \p12.95 & 0.096 $\pm$ 0.008 & --2.86 $\pm$  0.02 &  --5.44 $\pm$ 0.01 & --57.87 $\pm$  0.01 &   270.85 $\pm$  0.01 & 1.001 & 0.025 & 0.017  \\
          &   A &  \p12.95 & 0.102 $\pm$ 0.006 & --2.90 $\pm$  0.01 &  --5.72 $\pm$ 0.02 & --41.09 $\pm$  0.01 &    62.44 $\pm$  0.01 & 0.984 & 0.018 & 0.024  \\
          &   C & \p\p8.31 & 0.095 $\pm$ 0.007 & --2.80 $\pm$  0.02 &  --5.94 $\pm$ 0.04 & --21.86 $\pm$  0.01 &   --9.88 $\pm$  0.01 & 0.994 & 0.022 & 0.049  \\
          &   C & \p\p7.89 & 0.082 $\pm$ 0.006 & --2.93 $\pm$  0.02 &  --5.94 $\pm$ 0.06 & --21.89 $\pm$  0.01 &   --9.73 $\pm$  0.02 & 0.993 & 0.019 & 0.078  \\
          &   E & \p\p4.94 & 0.080 $\pm$ 0.007 & --2.51 $\pm$  0.02 &  --5.16 $\pm$ 0.01 &   31.43 $\pm$  0.01 &   308.78 $\pm$  0.00 & 0.953 & 0.024 & 0.012  \\
          &   E & \p\p4.52 & 0.084 $\pm$ 0.007 & --2.50 $\pm$  0.02 &  --5.17 $\pm$ 0.01 &   31.44 $\pm$  0.01 &   308.77 $\pm$  0.00 & 0.969 & 0.024 & 0.011  \\
          &   E & \p\p1.99 & 0.084 $\pm$ 0.004 & --2.53 $\pm$  0.01 &  --5.33 $\pm$ 0.01 &   29.22 $\pm$  0.00 &   299.72 $\pm$  0.01 & 0.996 & 0.012 & 0.013  \\
          &   C &  --11.07 & 0.100 $\pm$ 0.012 & --3.08 $\pm$  0.03 &  --5.55 $\pm$ 0.02 &  --2.08 $\pm$  0.01 &   --4.18 $\pm$  0.01 & 0.991 & 0.039 & 0.022  \\
          &   F &  --44.77 & 0.087 $\pm$ 0.007 & --2.94 $\pm$  0.02 &  --5.05 $\pm$ 0.02 & --61.71 $\pm$  0.01 &   196.77 $\pm$  0.01 & 0.991 & 0.022 & 0.020  \\
          &   D &  --46.04 & 0.083 $\pm$ 0.005 & --3.07 $\pm$  0.01 &  --5.98 $\pm$ 0.03 &  --9.00 $\pm$  0.00 & --299.53 $\pm$  0.01 & 0.978 & 0.016 & 0.033  \\
          &   D &  --46.46 & 0.081 $\pm$ 0.006 & --3.06 $\pm$  0.01 &  --6.03 $\pm$ 0.02 &  --8.98 $\pm$  0.01 & --299.52 $\pm$  0.01 & 0.975 & 0.018 & 0.020  \\
          &     &          &                   &                    &                    &                     &                      &       &       &        \\
    \Jtwo &   B &  \p12.95 & 0.102 $\pm$ 0.007 &  --2.95 $\pm$ 0.02 & --5.25 $\pm$  0.04 & --56.06 $\pm$  0.01 &   269.15 $\pm$  0.02 & 0.969 & 0.016 & 0.046  \\
          &   A &  \p12.95 & 0.111 $\pm$ 0.007 &  --2.99 $\pm$ 0.02 & --5.55 $\pm$  0.03 & --39.29 $\pm$  0.01 &    60.75 $\pm$  0.01 & 0.976 & 0.017 & 0.034  \\
          &   C & \p\p8.31 & 0.102 $\pm$ 0.005 &  --2.88 $\pm$ 0.01 & --5.77 $\pm$  0.06 & --20.06 $\pm$  0.00 &  --11.57 $\pm$  0.02 & 0.999 & 0.011 & 0.064  \\
          &   C & \p\p7.89 & 0.089 $\pm$ 0.006 &  --3.01 $\pm$ 0.02 & --5.77 $\pm$  0.07 & --20.08 $\pm$  0.01 &  --11.43 $\pm$  0.02 & 0.981 & 0.015 & 0.074  \\
          &   E & \p\p4.94 & 0.092 $\pm$ 0.006 &  --2.60 $\pm$ 0.01 & --4.98 $\pm$  0.04 &   33.24 $\pm$  0.01 &   307.09 $\pm$  0.01 & 0.969 & 0.013 & 0.041  \\
          &   E & \p\p4.52 & 0.098 $\pm$ 0.006 &  --2.59 $\pm$ 0.01 & --4.99 $\pm$  0.04 &   33.25 $\pm$  0.01 &   307.07 $\pm$  0.01 & 0.973 & 0.014 & 0.039  \\
          &   E & \p\p1.99 & 0.092 $\pm$ 0.008 &  --2.62 $\pm$ 0.02 & --5.14 $\pm$  0.05 &   31.03 $\pm$  0.01 &   298.03 $\pm$  0.02 & 0.981 & 0.021 & 0.049  \\
          &   C &  --11.07 & 0.103 $\pm$ 0.009 &  --3.17 $\pm$ 0.02 & --5.36 $\pm$  0.05 &  --0.27 $\pm$  0.01 &   --5.87 $\pm$  0.02 & 0.992 & 0.025 & 0.053  \\
          &   F &  --44.77 & 0.099 $\pm$ 0.006 &  --3.03 $\pm$ 0.01 & --4.88 $\pm$  0.03 & --59.91 $\pm$  0.01 &   195.08 $\pm$  0.01 & 0.974 & 0.013 & 0.035  \\
          &   D &  --46.04 & 0.090 $\pm$ 0.007 &  --3.16 $\pm$ 0.02 & --5.81 $\pm$  0.04 &  --7.19 $\pm$  0.01 & --301.23 $\pm$  0.01 & 0.973 & 0.018 & 0.038  \\
          &   D &  --46.46 & 0.089 $\pm$ 0.008 &  --3.15 $\pm$ 0.02 & --5.85 $\pm$  0.04 &  --7.18 $\pm$  0.01 & --301.21 $\pm$  0.01 & 0.993 & 0.021 & 0.039  \\
          &     &          &                   &                    &                    &                     &                      &       &       &        \\
 Combined &     &          &                   &                    &                    &                     &                      &       &       &        \\
          & all &          & 0.090 $\pm$ 0.006 &                    &                    &                     &                      & 0.970 & 0.026 & 0.051  \\

\enddata

\tablecomments{
Absolute proper motions are defined as $\mux = \mu_{\alpha \cos{\delta}}$ and $\muy = \mu_{\delta}$.
$\chi^2_\nu$ is reduced $\chi^2$ of post-fit residuals,  $\sigma_x$ and $\sigma_y$ are
error floor in $x$ and $y$, respectively.
}
\end{deluxetable}
\clearpage

\begin{deluxetable}{llccrr}
\tablecolumns{6}
\tablewidth{0pc}
\tablecaption{Best-fitting models for \hho\ maser kinematics in W49N}
\tablehead{
\multicolumn{3}{c}{Parameter} & \colhead{} & \multicolumn{2}{c}{Model} \\
\cline{1-3} \cline{5-6} \\
\colhead{} & \colhead{Type}  & \colhead{Unit} & \colhead{} & \colhead{A} & \colhead{B} }
\startdata
  Velocity &      $V_{0x}$ &   \kms &  & 101.34 $\pm$ 4.02 & 106.27 $\pm$ 4.28  \\
           &      $V_{0y}$ &   \kms &  &  61.30 $\pm$ 2.58 &  57.92 $\pm$ 3.13  \\
           &      $V_{0z}$ &   \kms &  &  10.73 $\pm$ 1.40 &  10.55 $\pm$ 1.28  \\
  Position &         $x_0$ & arcsec &  &   0.37 $\pm$ 0.08 &   0.23 $\pm$ 0.08  \\
           &         $y_0$ & arcsec &  &   0.06 $\pm$ 0.08 & --0.11 $\pm$ 0.13  \\
 Expansion & $V_{\rm exp}$ &   \kms &  &  16.77 $\pm$ 1.80 &  10.99 $\pm$ 1.33  \\
           &      $\gamma$ &        &  &   0.60 $\pm$ 0.06 &   0.81 $\pm$ 0.06  \\
  Rotation & $V_{\rm rot}$ &   \kms &  &          0        &   5.28 $\pm$ 0.80  \\
           &      $\theta$ & radian &  &          0        &   0.48 $\pm$ 0.72  \\
           &        $\phi$ & radian &  &          0        &   1.68 $\pm$ 0.30  \\
\enddata
\tablecomments{
Model A includes only radial motions. Model B includes radial 
and rotation motions.}
\label{tab:fit_g043}
\end{deluxetable}

\begin{deluxetable}{lccrrrrrr rrr } 
  \tabletypesize{\footnotesize} 
\tablecaption{Parallax and proper motion fits for \GVIII~\label{tab:g048_para_pm}}
\tablewidth{0pt} 
\tablehead{ 

\colhead{Background} & \colhead{Region} &  \colhead{\VLSR} & \colhead{Parallax} &    \colhead{\mux} &    \colhead{\muy} &   \colhead{\dx} &   \colhead{\dy} & \colhead{$\chi^2_\nu$} & \colhead{$\sigma_x$} & \colhead{$\sigma_y$} \\
    \colhead{source} &       \colhead{} & \colhead{(\kms)} &    \colhead{(mas)} & \colhead{(\masy)} & \colhead{(\masy)} & \colhead{(mas)} & \colhead{(mas)} &             \colhead{} &  \colhead{(mas)} &  \colhead{(mas)}

}
\startdata

  \Jthree &   A & 25.58 & 0.098 $\pm$ 0.003 & --3.11 $\pm$ 0.01 & --5.80 $\pm$ 0.02 &   2.81 $\pm$ 0.00 &  --4.68 $\pm$ 0.01 & 0.994 & 0.010 & 0.027 \\
          &   A & 25.16 & 0.097 $\pm$ 0.003 & --3.12 $\pm$ 0.01 & --5.77 $\pm$ 0.02 &   2.81 $\pm$ 0.00 &  --4.71 $\pm$ 0.01 & 0.996 & 0.009 & 0.026 \\
          &   A & 24.74 & 0.082 $\pm$ 0.003 & --3.12 $\pm$ 0.01 & --5.76 $\pm$ 0.07 &   2.82 $\pm$ 0.00 &  --4.79 $\pm$ 0.02 & 0.999 & 0.010 & 0.084 \\
          &   A & 24.31 & 0.086 $\pm$ 0.003 & --3.12 $\pm$ 0.01 & --5.77 $\pm$ 0.05 &   2.82 $\pm$ 0.00 &  --4.77 $\pm$ 0.02 & 0.997 & 0.009 & 0.063 \\
          &   B & 20.52 & 0.088 $\pm$ 0.004 & --2.67 $\pm$ 0.01 & --5.68 $\pm$ 0.02 & 131.04 $\pm$ 0.00 & --69.53 $\pm$ 0.01 & 0.996 & 0.011 & 0.029 \\
          &   B & 20.10 & 0.085 $\pm$ 0.003 & --2.67 $\pm$ 0.01 & --5.68 $\pm$ 0.02 & 131.03 $\pm$ 0.00 & --69.54 $\pm$ 0.01 & 0.988 & 0.010 & 0.028 \\
          &   B & 19.68 & 0.077 $\pm$ 0.003 & --2.69 $\pm$ 0.01 & --5.66 $\pm$ 0.02 & 131.01 $\pm$ 0.00 & --69.58 $\pm$ 0.01 & 0.938 & 0.009 & 0.028 \\
          &   B & 19.26 & 0.089 $\pm$ 0.004 & --2.72 $\pm$ 0.01 & --5.72 $\pm$ 0.02 & 130.04 $\pm$ 0.00 & --72.30 $\pm$ 0.01 & 1.009 & 0.013 & 0.026 \\
          &   B & 18.84 & 0.089 $\pm$ 0.004 & --2.73 $\pm$ 0.01 & --5.72 $\pm$ 0.02 & 130.05 $\pm$ 0.00 & --72.30 $\pm$ 0.01 & 1.002 & 0.013 & 0.027 \\
          &     &       &                   &                   &                   &                   &                    &       &       &       \\
  \Jseven &   A & 25.58 & 0.105 $\pm$ 0.004 & --3.18 $\pm$ 0.01 & --5.64 $\pm$ 0.02 &   2.98 $\pm$ 0.00 &    0.48 $\pm$ 0.01 & 0.994 & 0.013 & 0.028 \\
          &   A & 25.16 & 0.104 $\pm$ 0.004 & --3.19 $\pm$ 0.01 & --5.61 $\pm$ 0.02 &   2.98 $\pm$ 0.00 &    0.45 $\pm$ 0.01 & 0.998 & 0.012 & 0.028 \\
          &   A & 24.74 & 0.089 $\pm$ 0.004 & --3.19 $\pm$ 0.01 & --5.60 $\pm$ 0.06 &   3.00 $\pm$ 0.00 &    0.37 $\pm$ 0.02 & 0.999 & 0.012 & 0.077 \\
          &   A & 24.31 & 0.092 $\pm$ 0.004 & --3.18 $\pm$ 0.01 & --5.61 $\pm$ 0.05 &   2.99 $\pm$ 0.00 &    0.39 $\pm$ 0.02 & 0.997 & 0.012 & 0.057 \\
          &   B & 20.52 & 0.096 $\pm$ 0.004 & --2.74 $\pm$ 0.01 & --5.52 $\pm$ 0.01 & 131.21 $\pm$ 0.00 & --64.37 $\pm$ 0.01 & 0.993 & 0.013 & 0.017 \\
          &   B & 20.10 & 0.093 $\pm$ 0.004 & --2.74 $\pm$ 0.01 & --5.52 $\pm$ 0.01 & 131.21 $\pm$ 0.00 & --64.38 $\pm$ 0.00 & 0.979 & 0.013 & 0.015 \\
          &   B & 19.68 & 0.085 $\pm$ 0.004 & --2.76 $\pm$ 0.01 & --5.50 $\pm$ 0.02 & 131.19 $\pm$ 0.00 & --64.42 $\pm$ 0.01 & 0.910 & 0.012 & 0.018 \\
          &   B & 19.26 & 0.097 $\pm$ 0.004 & --2.79 $\pm$ 0.01 & --5.57 $\pm$ 0.01 & 130.21 $\pm$ 0.00 & --67.14 $\pm$ 0.00 & 0.995 & 0.014 & 0.015 \\
          &   B & 18.84 & 0.096 $\pm$ 0.004 & --2.80 $\pm$ 0.01 & --5.56 $\pm$ 0.01 & 130.22 $\pm$ 0.00 & --67.14 $\pm$ 0.00 & 0.998 & 0.013 & 0.015 \\
          &     &       &                   &                   &                   &                   &                    &       &       &       \\
   \Jfour &   A & 25.58 & 0.106 $\pm$ 0.005 & --3.28 $\pm$ 0.01 & --5.44 $\pm$ 0.04 &   2.68 $\pm$ 0.00 &    0.00 $\pm$ 0.01 & 0.996 & 0.015 & 0.043 \\
          &   A & 25.16 & 0.104 $\pm$ 0.005 & --3.28 $\pm$ 0.01 & --5.41 $\pm$ 0.04 &   2.68 $\pm$ 0.00 &  --0.03 $\pm$ 0.01 & 0.998 & 0.015 & 0.048 \\
          &   A & 24.74 & 0.089 $\pm$ 0.005 & --3.28 $\pm$ 0.01 & --5.40 $\pm$ 0.07 &   2.69 $\pm$ 0.00 &  --0.11 $\pm$ 0.03 & 0.998 & 0.015 & 0.089 \\
          &   A & 24.31 & 0.093 $\pm$ 0.005 & --3.28 $\pm$ 0.01 & --5.41 $\pm$ 0.06 &   2.69 $\pm$ 0.00 &  --0.09 $\pm$ 0.02 & 0.998 & 0.016 & 0.070 \\
          &   B & 20.52 & 0.096 $\pm$ 0.005 & --2.83 $\pm$ 0.01 & --5.32 $\pm$ 0.03 & 130.91 $\pm$ 0.00 & --64.84 $\pm$ 0.01 & 0.997 & 0.016 & 0.037 \\
          &   B & 20.10 & 0.093 $\pm$ 0.005 & --2.83 $\pm$ 0.01 & --5.32 $\pm$ 0.03 & 130.91 $\pm$ 0.00 & --64.86 $\pm$ 0.01 & 0.994 & 0.016 & 0.037 \\
          &   B & 19.68 & 0.087 $\pm$ 0.006 & --2.85 $\pm$ 0.02 & --5.30 $\pm$ 0.03 & 130.89 $\pm$ 0.01 & --64.90 $\pm$ 0.01 & 0.959 & 0.017 & 0.038 \\
          &   B & 19.26 & 0.097 $\pm$ 0.005 & --2.89 $\pm$ 0.01 & --5.36 $\pm$ 0.03 & 129.91 $\pm$ 0.00 & --67.62 $\pm$ 0.01 & 0.997 & 0.015 & 0.037 \\
          &   B & 18.84 & 0.097 $\pm$ 0.005 & --2.89 $\pm$ 0.01 & --5.36 $\pm$ 0.03 & 129.92 $\pm$ 0.00 & --67.62 $\pm$ 0.01 & 0.999 & 0.015 & 0.036 \\
          &     &       &                   &                   &                   &                   &                    &       &       &       \\
 Combined &     &       &                   &                   &                   &                   &                    &       &       &       \\
          & all &       & 0.093 $\pm$ 0.005 &                   &                   &                   &                    & 0.990 & 0.029 & 0.069 \\

\enddata
\tablecomments{
Absolute proper motions are defined as $\mux = \mu_{\alpha \cos{\delta}}$ and $\muy = \mu_{\delta}$.
$\chi^2_\nu$ is reduced $\chi^2$ of post-fit residuals,  $\sigma_x$ and $\sigma_y$ are
error floor in $x$ and $y$, respectively.
}

\end{deluxetable}
\clearpage

\begin{deluxetable}{llccrr}
\tablecolumns{6}
\tablewidth{0pc}
\tablecaption{Best-fitting models for \hho\ maser kinematics in \GVIII}
\tablehead{
\multicolumn{3}{c}{Parameter} & \colhead{} & \multicolumn{2}{c}{Model} \\
\cline{1-3} \cline{5-6} \\
\colhead{} & \colhead{Type}  & \colhead{Unit} & \colhead{} & \colhead{A} & \colhead{B} }
\startdata
  Velocity &      $V_{0x}$ &   \kms &  &  15.51  $\pm$  3.96 &  25.00  $\pm$   4.44\\
           &      $V_{0y}$ &   \kms &  &   5.44  $\pm$  3.83 &   6.62  $\pm$   3.65\\
           &      $V_{0z}$ &   \kms &  &  20.65  $\pm$  2.64 &  19.24  $\pm$   2.73\\
  Position &         $x_0$ & arcsec &  &   0.11  $\pm$  0.09 & --0.06  $\pm$   0.11\\
           &         $y_0$ & arcsec &  & --0.05  $\pm$  0.09 &   0.01  $\pm$   0.09\\
 Expansion & $V_{\rm exp}$ &   \kms &  &   7.78  $\pm$  2.91 &   9.25  $\pm$   3.30\\
           &      $\gamma$ &        &  & --0.26  $\pm$  0.35 & --0.23  $\pm$   0.41\\
  Rotation & $V_{\rm rot}$ &   \kms &  &           0         &   7.44  $\pm$   8.45\\
           &      $\theta$ & radian &  &           0         &   1.48  $\pm$   1.76\\
           &        $\phi$ & radian &  &           0         & --1.46  $\pm$   1.59\\
\enddata
\tablecomments {
Model A includes only radial motions. Model B includes radial 
and rotation motions.}
\label{tab:fit_g048}
\end{deluxetable}

\begin{deluxetable}{ccccccccc}
\tablecolumns{9}
\tablewidth{0pc}
\tablecaption{Parallaxes and proper motions of \hho\ maser sources located in Inner Perseus Arm}
\tablehead{
\colhead{Source} & \colhead{Parallax} & \colhead{Distance} & \colhead{\mux}    & \colhead{\muy}    & \colhead{\VLSR} & \colhead{\Us}    & \colhead{\Vs}    & \colhead{\Ws} \\
\colhead{name}   & \colhead{(mas)}    & \colhead{(kpc)}    & \colhead{(\masy)} & \colhead{(\masy)} & \colhead{(\kms)} & \colhead{(\kms)} & \colhead{(\kms)} & \colhead{(\kms)}}
\startdata
\GIII  & 0.090 $\pm$ 0.006 & $11.11^{+0.79}_{-0.69}$ & --2.48 $\pm$ 0.15 & --5.27 $\pm$ 0.13 & 11 $\pm$ 5 & --17 $\pm$  11 & --23 $\pm$  16 & --5 $\pm$  8\\
\GVIII & 0.093 $\pm$ 0.005 & $10.75^{+0.61}_{-0.55}$ & --2.89 $\pm$ 0.13 & --5.50 $\pm$ 0.13 & 18 $\pm$ 5 &\p--4 $\pm$  11 & \p13 $\pm$  14 & \p6 $\pm$  7\\
\enddata
\tablecomments{
Column 3 lists the parallax distance. Column 4 to 6 list absolute proper
motion in eastward and northward direction and \VLSR, respectively.
Columns 7 to 9 list peculiar motion components, where \Us, \Vs, \Ws\ are
directed toward the Galactic center, in the direction of Galactic
rotation and toward the North Galactic Pole (NGP), respectively. The
peculiar motions were estimated using the Galactic parameters from
\citet{2011AN....332..461B} and solar motion parameters from
\citet{2010MNRAS.403.1829S}.
}
\label{tab:ppm}
\end{deluxetable}

\clearpage

\end{document}